\documentclass[a4paper,fleqn,usenatbib]{mnras}

\usepackage{newtxtext,newtxmath}
\usepackage[T1]{fontenc}
\usepackage{ae,aecompl}
\usepackage{pdflscape}
\usepackage{rotating}

\usepackage{graphicx}	% Including figure files
\usepackage{amsmath}	% Advanced maths commands
\usepackage{amssymb}	% Extra maths symbols

\title[A chemical study of M67 BSSs]{A chemical study of M67 candidate blue stragglers and evolved blue stragglers observed with APOGEE DR14}

\author[Bertelli Motta et al.]{
	Clio Bertelli Motta,$^{1}$\thanks{E-mail: cbertelli@ari.uni-heidelberg.de}
	Anna Pasquali,$^{1}$
	Elisabetta Caffau,$^{2}$
	and Eva K. Grebel$^{1}$
	\\
	$^{1}$Astronomisches Rechen-Institut, Zentrum f\"ur Astronomie der Universit\"at Heidelberg, M\"onchhofstr. 12-14, 69120 Heidelberg, Germany\\
	$^{2}$GEPI, Observatoire de Paris, Universit\'e PSL, CNRS, Place Jules Janssen, 92190 Meudon, France}

\date{Accepted 2018 August 06. Received 2018 August 03; in original form 2018 May 07}

\pubyear{2018}

\begin{document}
\label{firstpage}
\pagerange{\pageref{firstpage}--\pageref{lastpage}}
\maketitle

\begin{abstract}
Within the variety of objects populating stellar clusters, blue straggler stars (BSSs) are among the most puzzling ones. BSSs are commonly found in globular clusters, but they are also known to populate open clusters of the Milky Way. Two main theoretical scenarios (collisions and mass transfer) have been suggested to explain their formation, although finding observational evidence in support of either scenario represents a challenging task. Among the APOGEE observations of the old open cluster M67, we found 8 BSS candidates known from the literature and two known evolved BSSs. We carried out a chemical analysis of 3 BSS candidates and of the 2 evolved BSSs out of the sample and found that the BSS candidates have surface abundances similar to those of stars on the main-sequence turn-off of M67. Especially the absence of any anomaly in their carbon abundances seems to support a collisional formation scenario for these stars. Furthermore, we note that the abundances of the evolved BSSs S1040 and S1237 are consistent with the abundances of the red clump stars of M67.  In particular, they show a depletion in carbon by $\sim0.25$ dex, which could be either interpreted as the signature of mass transfer or as the product of stellar evolutionary processes. Finally, we summarise the properties of the individual BSS stars observed by APOGEE, as derived from their APOGEE spectra and/or from information available in the literature.
\end{abstract}


%%%%%%%%%%%%%%%%%%%%%%%%%%%%%%%%%%%%%%%%%%%%%%%%%%

%%%%%%%%%%%%%%%%% BODY OF PAPER %%%%%%%%%%%%%%%%%%

\section{Introduction}
\label{sec:intr}
Among the variety of objects that populate stellar clusters, blue straggler stars (BSSs) are surely among those still presenting many riddles to astronomers. They are more luminous and/or bluer than the cluster turn-off (TO), but their distribution in the colour-magnitude diagram (CMD) cannot be explained by a younger single stellar population. BSSs were first discovered by \citet{sandage1953} in the CMD of the globular cluster M3. Shortly after, several old open clusters, e.g. NGC 7789 \citep{burbidge1958}, M67 \citep{johnson1955}, NGC 188 \citep{sandage1962}, and NGC 6791 \citep{kinman1965}, were found or at least suspected to host a BSS population (for a review of the historical discoveries of BSSs see \citealt{cannon2015}). Nowadays, BSSs are known to populate a large number of open and globular clusters (for a catalogue, see \citealt{ahumada2007} and \citealt{fusipecci1993}, respectively). 

In order to appear 'younger' than the age of their hosting cluster, BSSs must have undergone processes that led to an increase in mass, e.g. through interactions with other stars, thus prolonging their life on the main sequence (MS).
There are several BSS formation scenarios predicted by theoretical studies. The first is mass transfer or mergers in close binary systems \citep{mccrea1964,strom1970}: when one of the stars in a close binary system fills its Roche lobe (e.g. after expansion due to its evolution along the subgiant branch, hereafter SGB), material from this star is accreted onto its companion, thus increasing its mass and extending its hydrogen burning phase for a longer time than its initial core mass would have allowed. Under some circumstances, the two stars might even merge. Recently, it has been suggested that BSSs could form via the merger of binaries in hierarchical triple systems caused by Kozai cycles and tidal friction \citep{perets2009,naoz2014}. In this scenario a close binary system, also called inner binary, acts as a companion to a third star. The dynamical interactions within the hierarchical triple system lead to oscillations in the eccentricity and inclination of the inner close binary (Kozai oscillations or Kozai cycles, first proposed in \citealt{kozai1962}), which might then merge into a BSS. This scenario would, among other things, explain the presence of BSSs in long-period, high-eccentricity binaries.
The second scenario involves dynamically induced stellar collisions between single stars or multiple systems that lead to a merger and thus to the formation of  a BSS \citep{hills1976}.

If BSSs form through mass transfer, the material that is accreted last comes from more inner layers of the donor, where e.g. the CNO cycle was active. We thus expect the BSS surface abundances of elements such as C, N, and O to be different from those of the parental cloud and consequently of the un-evolved cluster members (see, e.g., \citealt{sarna1996}).
If, instead, BSSs form through a merger following the dynamical collision of two stars, simulations predict only a small amount of mixing between the stellar interior and the outer layers. For this reason, surface abundances in collisional BSSs are expected to be the same as in MS or TO stars (see, e.g., \citealt{lombardi1995}).
Thus, investigating the surface chemical composition of  BSSs can hint at the circumstances  under which they formed.

The measurement of surface abundances in BSSs belonging to globular clusters is difficult because of their faintness, but in recent years several studies have been carried out trying to determine the chemical composition of these objects and thus to constrain their formation history. \citet{ferraro2006} studied the chemical composition of BSSs in 47 Tuc and found a substantial fraction of stars depleted in C and O. Nevertheless, 47 Tuc is the only cluster in which such a high number of depleted BSSs has been observed. In M4, M30, and $\omega$ Centauri a maximum of $1-2$ stars with a significant CO signature have been detected \citep[see, e.g.,][]{lovisi2010,lovisi2013,mucciarelli2014}. This leads to the conclusion that either the mass transfer formation channel is not very efficient in globular clusters or the surface C and O depletion expected in this scenario is only a temporary feature \citep[][and references therein]{ferraro2015}.

The old open cluster M67 ($\sim4$ Gyr) is known to host a relatively large population of BSSs  and has been one of the first clusters in which these peculiar objects were discovered \citep{johnson1955}. The M67 BSSs have since then been extensively studied observationally both from a photometric \citep[e.g.][]{gilliland1991,vandenberg2002,stassun2002,sandquist2003a,sandquist2003b,pribulla2008} and spectroscopic point of view \citep[e.g.][]{mathieu1986,mathieu1990,mathys1991,latham1996,shetrone2000,liu2008,geller2015}, and also theoretically \citep[e.g.][]{tian2006,hurley2005,hurley2001}. Although dynamical collisions are expected to be a more relevant formation channel in denser environments, N-body simulations have shown that stellar dynamics plays an important role in the formation of the BSSs population found in M67. Nevertheless, the BSSs formed in the simulations through dynamical processes alone are not enough, and a large number of primordial close binaries has to be invoked to match the observed BSSs \citep{hurley2005,hurley2001}. Both formation  scenarios (mass transfer and collisions) thus seem to concur in the formation of BSSs in M67, at least from the point of view of numerical simulations.

Within the field of M67, 24 stars are known as candidate BSSs from \citet{deng1999} (see Table~\ref{tab:candidates}). These stars have been subject to numerous studies over the years, both spectroscopically \citep[e.g.][]{liu2008,geller2015} and photometrically \citep[e.g.][]{sandquist2003b,pribulla2008}. In particular, \citet{liu2008} performed a spectroscopic study of 19 stars from the list of BSS candidates and derived their $T_\mathrm{eff}$ and $\log g$. \citet{sandquist2003b} studied the light curves of 20 of these stars, indicating several variable stars among them that will be discussed later in this paper. \citet{pribulla2008} investigated the variability of all the 24 BSS candidates with the MOST (\textit{Microvariability and Oscillations of STars}) satellite (see Table~\ref{tab:candidates}).  More recently, \citet{geller2015} included all 24 objects in a larger list of targets for their study of the kinematics of M67. For each of them they determined their membership probability and binarity. Furthermore, they excluded some of these stars from the list of BSSs based either on their kinematic properties or on their position on the CMD. In Table~\ref{tab:candidates} we list for each BSS candidate from \citet{deng1999} whether it is considered a member and a blue straggler by \citet{geller2015} and a variable by \citet{sandquist2003b} and \citet{pribulla2008}.

Furthermore, M67 hosts other peculiar objects such as red straggler and yellow straggler stars (YSS, or yellow giants). The first ones, also known as sub-subgiants, are objects less luminous than the SGB and redder than the MS, but consistent with the kinematics of the cluster. \citet{mathieu2003} found two of these objects in the field of M67, S1113 and S1063 (naming from \citealt{sanders1977}), which revealed themselves as binaries composed of a SGB star and a main-sequence star and whose origin and low luminosity are still unexplained. YSS are objects more luminous than the SGB, but bluer than the red giant branch (RGB). Two of the YSS present in M67 are known to be evolved BSSs, S1040 and S1237. The first has been studied in detail by \citet{landsman1997}, who found it to be composed by a red giant and a He-core white dwarf companion. S1237 was subject to a study carried out by \citet{leiner2016} based on  Kepler asteroseismic data \citep{howell2014}. The authors derived the asteroseismic mass and radius of the primary companion of the binary system, that was found to be much more massive ($\sim2.9\,M_{\odot}$) than the M67 red clump (RC). A third star, S1072, is listed in the literature as a candidate BSS but is actually redder than the TO, and is thus thought to be evolving along its SGB. In \citet{geller2015} this star is listed as a yellow giant (see Table~\ref{tab:candidates}).

Previous investigations of the chemical composition (Fe, Li, C, N, Na, Mg, Ca, Ni, Ba) of 7 BSSs in M67 (\citealt{shetrone2000,mathys1991})  found that the surface abundances of these stars are consistent with those of TO stars, which would suggest a collisional formation scenario. In the present study, we perform an independent chemical analysis of the APOGEE infrared spectra of 3 candidate BSSs and 2 evolved BSSs in the field of M67 in order to understand whether they can give us clues about the BSSs formation mechanisms in M67. In Section~\ref{sec:method}, we present the data and the tools  used in our investigation. The chemical analysis of the 5 stars under study and its results are discussed in Section~\ref{sec:res}. In Section~\ref{sec:disc}, we extend the discussion to include other properties such as the spatial position and the rotational velocity of all BSS candidates and evolved BSSs observed with APOGEE DR14 in the light of our present understanding of BSSs. Furthermore, we review the information available about these stars putting our results on their chemical composition, when available, in a more general context that includes the stellar parameters derived within APOGEE DR14 and information gathered from the literature. In Section~\ref{sec:concl}, we then summarise the main conclusions of the present study.

\section{Data and Method}
\label{sec:method}

Out of the 24 BSS candidates from \citet{deng1999}, 12 are included in the fourteenth data release of the Apache Point Observatory Galactic Evolution Experiment \citep[hereafter APOGEE DR14, see][]{majewski2017,sdss14}. Four of them (S977, S1434, S1066, and S968) have the ASPCAP flag set to STAR\_BAD \footnote{"BAD overall for star: set if any of TEFF, LOGG, CHI2, COLORTE, ROTATION, SN error are set, or any parameter is near grid edge (GRIDEDGE\_BAD is set in any PARAMFLAG", see http://www.sdss.org/dr14/algorithms/bitmasks/\#APOGEE\_ASPCAPFLAG} and we therefore do not take these stars into further consideration (see Table~\ref{tab:candidates}). The other stars have, if any, only warning flags. In addition, APOGEE DR14 also observed the two known evolved BSSs of M67, S1040 and S1237 (see Table~\ref{tab:candidates}).  

\begin{figure}
	\includegraphics[width=\columnwidth]{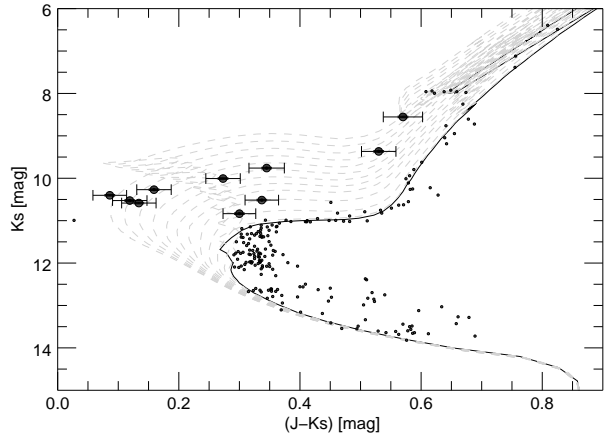}
	\caption{CMD of M67 where the kinematic members of the cluster are represented as small black dot and the 8 BSS candidates as well as the two evolved BSS considered in this studied are plotted as large black dots. We also show a sequence of PARSEC isochrones \citep{bressan2012} ranging from 1 to 4 Gyr, with $E(B-V)=0.023$ mag and $(m-M)_0=9.64$ mag \citep{bellini2010}. The 4 Gyr isochrone is plotted as a black solid line, the others as dashed grey lines.}
	\label{fig:iso_seq}
\end{figure}

We use the effective temperatures, gravities, and microturbulences (when available) derived with the APOGEE Stellar Parameters and Chemical Abundances Pipeline (ASPCAP) \citep[see][]{garciaperez2015,holtzman2015} as an input for our independent chemical analysis performed with MyGIsFOS \citep[for details see][]{sbordone2014}. The code MyGIsFOS uses a pre-computed synthetic stellar grid to determine stellar parameters and abundances. Although the code allows for the determination of stellar parameters such as $T_\mathrm{eff}, \log g, V_\mathrm{micro}$, and [$\alpha$/Fe], in order to obtain, e.g., $\log g$, FeII lines are required, which are not present in the spectra under study. We thus keep the stellar parameters ($T_\mathrm{eff}, \log g, V_\mathrm{micro}$ ) available in ASPCAP DR 14 and [$\alpha$/Fe]=0. Since for dwarf stars no calibrated $\log g$ values are available in APOGEE DR14, we use $\log g$ values derived comparing the position of the stars on the CMD with the closest isochrone (see Fig.~\ref{fig:iso_seq}). MyGIsFOS then computes the element abundances fitting the different line profiles with the synthetic spectra along the [Fe/H]--dimension of the grid. We choose not to use the abundances derived from ASPCAP in order to have a better understanding of the systematics involved in the abundance determination and in order to be able to carry out tests, e.g., on the dependence of the obtained abundances on $T_\mathrm{eff},\log g$, and the persistence (i.e., residual flux from previous exposures) present in some of the APOGEE spectra \citep{holtzman2015}.

\begin{table*}
	\caption{The 24 candidate BSSs from \citet{deng1999}, as well as the two evolved BSS S1237 and S1040 are listed together with their membership probability and confirmed BSS or yellow giant (YG) nature from \citet{geller2015} (G15), and their variability as detected by \citet{sandquist2003b} (S\&S03) and  \citet{pribulla2008} (P08). BLM: binary likely member, BM: binary member, SM: single member, SN: single non-member, DS: $\delta$ Scuti variable, EB: eclipsing binary, Poss. LA: possible low amplitude. In the last five columns we indicate which stars are present in APOGEE DR14, which have the flag STAR\_BAD, which are fast rotators ($v\sin i>50$ km s$^{-1}$), which are too hot to be analysed ($T_\mathrm{eff}>7000$ K), and which have been analysed in the present work.}
	\vspace{0.5cm}
	\begin{tabular}{l|c|c|c|c|c|c|c|c|c}
		\hline
		\multicolumn{1}{c|}{Name} &
		\multicolumn{1}{c|}{Memb. G15} &
		\multicolumn{1}{c|}{G15} &
		\multicolumn{1}{c|}{Var. S\&S03} &
		\multicolumn{1}{c|}{Var. P08} &
		\multicolumn{1}{c}{APOGEE DR14} &
		\multicolumn{1}{c}{STAR\_BAD} &
		\multicolumn{1}{c}{Fast Rot.} &
		\multicolumn{1}{c}{Hot} & 
		\multicolumn{1}{c}{This work}\\
		\hline
		\hline
		S977	&	(BL)M		&	BSS	& N &	--		&	X	& X & -- & -- & --\\
		S1434	&	(BL)M		&	BSS	& -- & --	&	X & X & -- & -- & --\\
		S1066	&	(BL)M		&	BSS	&	N &	--	&	X & X & -- & -- & --\\
		S1267	&	BM		&	BSS	&	N & --		&	X & -- & X & X & -- \\
		S1284	&	BM		&	BSS	& DS &	DS		&	X & -- & X & X & -- \\
		S1263	&	SM		&	BSS	&	Poss. LA & 	--	&	X & -- & -- & X & --\\
		S968	&	SM		&	BSS	&	Poss. LA, Am &	--	&	X& X & -- & -- & --\\
		S975	&	BM		&	BSS	&	Faint comp. & --	&	--	& --& -- & -- & --	\\
		S1082	&	BM		&	BSS	&	RS CVn var. & EB	&		X & -- & --& X & --\\
		S752	&	BM		&	BSS	&	Am, poss. flare &	--	&	X & -- & X & -- & --\\
		S1072	&	BM		&	YG	&	N &	--	&	X & -- & -- & --& X \\
		S1280	&	(BL)M		&	BSS	& DS &	DS	&   -- & -- & -- & -- & --\\
		S997	&	BM		&	BSS		& N &	--	& -- & -- & -- & -- & --\\
		S1195	&	BM		&	BSS		&	N &--	& -- & -- & --& -- & --\\
		S792	&	SM		&	--	&	N & 	--	&	X& -- & -- & -- & X \\
		S277	&	BM		& --		&	-- & --	& --  & -- & --& -- & --\\
		S2226	&	SM		&	BSS		& -- &	--	& -- & -- & -- & -- & --\\
		S1273	&	SM		&	--		&	N & --	& -- & -- & --& -- & --\\
		S984	&	SM		&	--		&	N & --	&	X & -- & -- & -- & X\\
		S1005	&	BM		&	--		&	N & --	& -- & -- & -- & -- & --\\
		S751	&	SM		&	--		&	N & --	& -- & -- & -- & -- &--\\
		S1036	&	(BL)M		&	--	&W UMa var. &	EB	&  -- & -- & --& -- & --\\
		S145	&	SN		&	--		&	-- & --	& --  & --& --& -- & --\\
		S2204	&	SM		&	--		&	N &--	& -- & -- &--& -- & --\\
		\hline
		S1237 & BM & YG &  -- & -- & X & -- & --& -- &X \\
		S1040 & BM & YG & Brightness var. & -- & X & --& --& -- & X\\
		\hline
		
	\end{tabular}
	\label{tab:candidates}
\end{table*}

\section{Analysis and Results}
\label{sec:res}

\begin{figure}
	\includegraphics[width=\columnwidth]{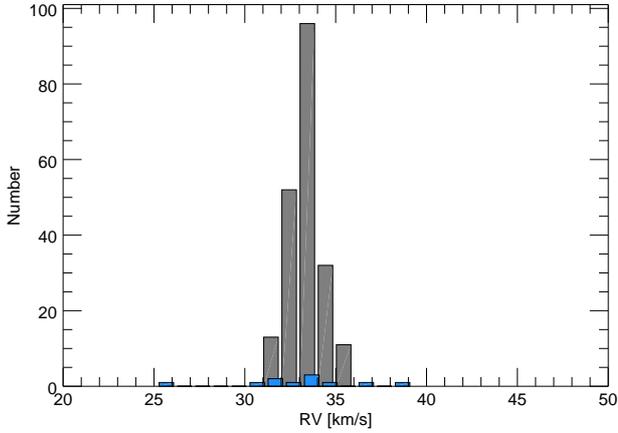}
	\caption{The grey histogram represents the APOGEE DR14 radial velocities of stars considered members based on their kinematic properties. The blue histogram shows the radial velocities obtained in APOGEE DR14 for the 8 BSS candidates and the 2 evolved BSSs.}
	\label{fig:hist}
\end{figure}

\begin{figure}
	\includegraphics[width=\columnwidth]{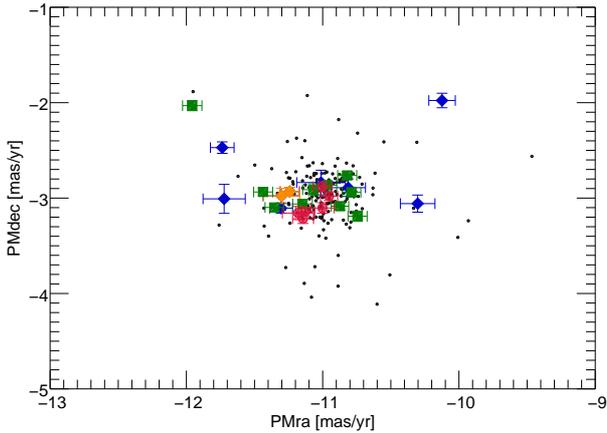}
	\caption{Proper motion in RA and Dec from Gaia DR2 of our sample of stars. The small black dots represent the stars selected through our membership analysis. The BSS candidates known from the literature and found in APOGEE DR14 are plotted as blue diamonds, while the two evolved BSSs are shown as orange diamonds. The two control samples of stars on the TO and on the RC are shown as green squares and red dots, respectively. See text for details.}
	\label{fig:pm}
\end{figure}

\begin{figure}
	\includegraphics[width=\columnwidth]{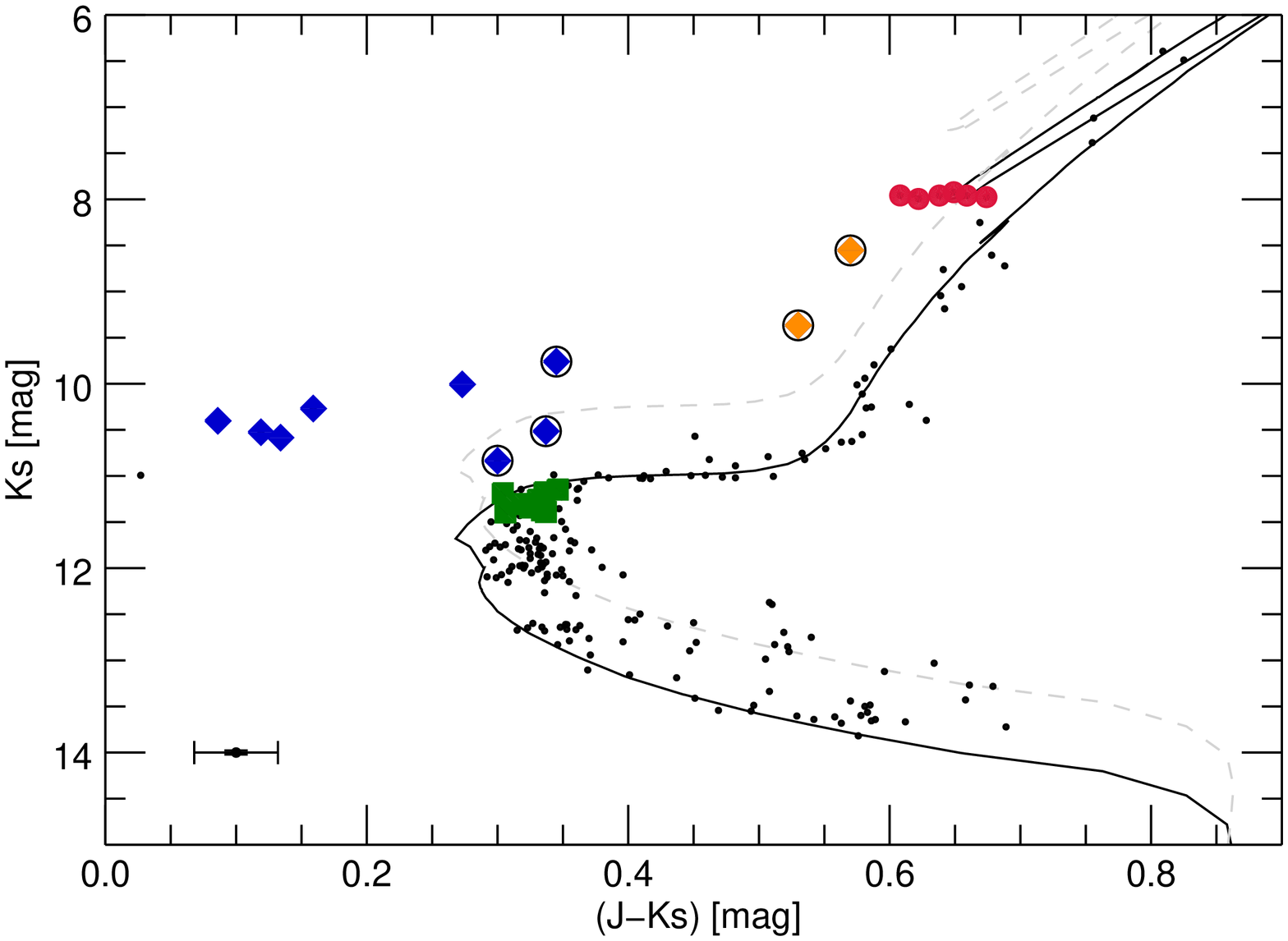}
	\caption{The plot shows the position on the CMD of the stars involved in our study. The colours are the same as in Fig.~\ref{fig:pm}. The stars for which we obtained chemical abundances are circled in black. We show a 4 Gyr PARSEC isochrone as a solid black line and the position of the equal-mass binaries as a dashed line ($E(B-V)=0.023$ mag and $(m-M)_0=9.64$ mag from \citealt{bellini2010}). At the bottom left of the plot are the mean photometric errors for the sample of M67 kinematics members.}
	\label{fig:iso}
\end{figure}

We first compare the sample of BSS candidates from \citet{deng1999} considered in our study with the list of M67 members obtained performing a membership analysis of the stars observed by APOGEE DR14 within $\sim1\deg$ (the outer radius of the cluster, according to \citealt{kharchenko2013}) from the centre of  M67 based only on their Gaia DR2 proper motions (see, e.g., \citealt{gaia2016,gaia2018b}) and APOGEE DR14 radial velocities, as in the first 4 steps of the membership analysis described in \citet{bertelli2017}.
In Fig.~\ref{fig:hist} we show the APOGEE DR14  radial velocity (RV) distribution of the BSS candidates and YSS considered in this work  in comparison to the RV distribution found for M67 after our membership analysis. Three BSS candidates lie outside of the cluster distribution obtained by our membership analysis and would have been eliminated by our selection. The RVs obtained by APOGEE for the BSS candidates are also in some cases significantly different from the ones derived by \citet{geller2015} (see Table~\ref{tab:rv}), and we suggest that the inconsistent velocities of these stars might be due to variations caused by the position of the star on its putative binary orbit at the time of observation. Nevertheless, the proper motions that we retrieve from the Gaia DR2 catalogue for the BSS candidates are consistent with the mean proper motion of M67 (see Fig.~\ref{fig:pm} and Table~\ref{tab:star_param}), except for S1082 that does not have an entry in Gaia DR2.  In our study, we also consider two objects located above the SGB of M67 and known from the literature as evolved BSSs, S1040 \citep{landsman1997} and S1237 \citep{leiner2016}. The complete sample, containing in total 10 stars, is displayed in Fig.~\ref{fig:iso}, where the 8 candidate BSSs are shown as blue diamonds, and the evolved BSSs as orange diamonds.

\begin{table}
	\caption{The rotation velocities and radial velocities of the 10 stars of our sample derived by APOGEE DR14 compared to the radial velocities from \citet{geller2015}.}
	\vspace{0.5cm}
	\begin{tabular}{l|c|c|c|c|c}
		\hline
		\multicolumn{1}{c|}{Name} &
		\multicolumn{1}{c|}{$v\sin i$} &
		\multicolumn{1}{c|}{AP RV} &
		\multicolumn{1}{c|}{AP RV\_e} &
		\multicolumn{1}{c|}{G15 RV} &
		\multicolumn{1}{c}{G15 RV\_e} \\
		\multicolumn{1}{c|}{} &
		\multicolumn{1}{c|}{[km s$^{-1}$]} &
		\multicolumn{1}{c|}{[km s$^{-1}$]} &
		\multicolumn{1}{c|}{[km s$^{-1}$]} &
		\multicolumn{1}{c|}{[km s$^{-1}$]} &
		\multicolumn{1}{c}{[km s$^{-1}$]} \\
		\hline
		\hline
		S792 & 3.63 & 34.08 & 0.01 & 33.49 & 0.1\\
		S752 & 76.26& 25.68 & 0.14 & 31.29 & 0.35\\
		S984 & 8.00 & 33.49 & 0.02 & 31.89 & 0.19\\
		S1072 & 13.88 & 34.88 & 0.02 & 32.72 & 0.13\\
		S1040 & 6.09& 31.69 & 0.01 & 33.01 & 0.08\\
		S1263 & 25.56& 32.67 & 0.07 & 32.22 & 0.07\\
		S1284 & 79.45 & 34.42 & 0.15 & 31.92 & 0.37\\
		S1267 & 60.90& 38.71 & 0.44 & 33.76 & 0.23\\
		S1237 & -- & 37.34 & 0.003 & 33.58 & 0.06\\
		S1082 & 13.99& 34.42 &0.04&33.68 &0.09\\
		\hline\end{tabular}
	
	\label{tab:rv}
\end{table}

\begin{table*}
	\caption{Proper motions from the Gaia DR2 catalogue \citep{gaia2016,gaia2018b}, except for S1082, which was not observed by Gaia, and 2MASS photometry \citep{skrutskie2006} for all the stars discussed in this work. }
	\vspace{0.5cm}
	
		\begin{tabular}{l|c|c|c|c|c|c|c|c|c|c}
			\hline
			\multicolumn{1}{c|}{Name} &
			\multicolumn{1}{c|}{PM\_ra} &
			\multicolumn{1}{c|}{PM\_dec} &
			\multicolumn{1}{c|}{e\_PM\_ra}&
			\multicolumn{1}{c|}{e\_PM\_dec} &
			\multicolumn{1}{c|}{Jmag} &
			\multicolumn{1}{c|}{e\_Jmag} &
			\multicolumn{1}{c}{Hmag} &
			\multicolumn{1}{c|}{e\_Hmag} &
			\multicolumn{1}{c|}{Ksmag} &
			\multicolumn{1}{c|}{e\_Ksmag} \\
			
			\multicolumn{1}{c|}{} &
			\multicolumn{1}{c|}{[mas yr$^{-1}$]} &
			\multicolumn{1}{c|}{[mas yr$^{-1}$]} &
			\multicolumn{1}{c|}{[mas yr$^{-1}$]}&
			\multicolumn{1}{c|}{[mas yr$^{-1}$]} &
			\multicolumn{1}{c|}{[mag]} &
			\multicolumn{1}{c|}{[mag]} &
			\multicolumn{1}{c}{[mag]} &
			\multicolumn{1}{c|}{[mag]} &
			\multicolumn{1}{c|}{[mag]} &
			\multicolumn{1}{c|}{[mag]} \\
			\hline
			\hline
			\multicolumn{11}{c}{BS stars}\\
			\hline   
			
			S977&   -11.577 & -2.352 & 0.145 & 0.113  & 10.137& 0.021 & 10.224& 0.022&  10.228& 0.018\\ 
			S1434&  -11.091 & -2.972 & 0.116 & 0.072 & 10.412& 0.027 & 10.427&  0.030&  10.406&  0.020\\  
			S1066&  -10.250 & -3.137 & 0.116 & 0.086 & 10.747& 0.022 & 10.733& 0.020 &  10.702& 0.020 \\ 
			S1267&   -10.303 & -3.058 & 0.127 & 0.090   & 10.488& 0.022 & 10.427& 0.022&  10.402& 0.017\\ 
			S1284&   -10.813 & -2.891 & 0.127 & 0.092  & 10.428& 0.022 & 10.318& 0.022&  10.269& 0.018\\ 
			S1263& -11.305 & -3.105 & 0.084 & 0.054 & 10.645& 0.022 & 10.541& 0.020 &  10.526& 0.018\\ 
			S968&   -11.374 & -2.692 & 0.115 & 0.075 & 11.020 & 0.021 & 11.011& 0.020 &  10.993& 0.018\\ 
			S1082& --& --& --& --& 10.280 & 0.022 & 10.080 & 0.022& 10.007 & 0.018\\ 
			S752&   -11.723 & -3.008 & 0.155 & 0.152 & 10.719& 0.022 & 10.621& 0.020 &  10.585& 0.018\\ 
			S1072&   -10.124 & -1.977 & 0.097 & 0.076  & 10.105& 0.023 & 9.816 & 0.022&  9.760  & 0.018\\ 
			S792&  -11.012 & -2.837 & 0.178 & 0.129   & 10.852& 0.021 & 10.586& 0.020 &  10.515& 0.018\\ 
			S984&  -11.737 & -2.471 & 0.087 & 0.060 & 11.135& 0.021 & 10.888& 0.020 &  10.835& 0.017\\ 
			S1040&  -11.302 & -2.977 & 0.135 & 0.080& 9.897 & 0.022 & 9.482 & 0.020 &  9.367 & 0.018\\ 
			S1237&  -11.242 & -2.932 & 0.070 & 0.045  & 9.124 & 0.027 & 8.661 & 0.026&  8.554 & 0.018\\ 
			\hline
			\multicolumn{11}{c}{TO stars}\\
			\hline   
			S598&  -10.789 & -2.940 & 0.069 & 0.047 &  11.513& 0.020 &  11.244& 0.023&  11.177& 0.021\\ 
			S815& -10.744 & -3.190 & 0.070 & 0.054 &  11.706& 0.022&  11.435& 0.019&  11.372& 0.017\\ 
			S756& -11.068 & -2.912 & 0.079 & 0.062 &  11.491& 0.022&  11.220 & 0.020 &  11.187& 0.020 \\ 
			S1076&  -11.957 & -2.031 & 0.071 & 0.053&  11.728& 0.022&  11.453& 0.020 &  11.391& 0.020 \\ 
			S1083& -10.873 & -3.085 & 0.064 & 0.043&  11.667& 0.022&  11.382& 0.022&  11.342& 0.020 \\ 
			S1310& -11.148 & -3.062 & 0.072 & 0.046 &  11.703& 0.021&  11.466& 0.022&  11.397& 0.019\\ 
			S1268& -11.354 & -3.098 & 0.069 & 0.043 &  11.494& 0.022&  11.196& 0.022&  11.148& 0.020 \\ 
			S1456& -11.437 & -2.936 & 0.070 & 0.043&  11.625& 0.022&  11.390 & 0.023&  11.305& 0.020 \\ 
			S1429& -10.820 & -2.762 & 0.070 & 0.043&  11.634& 0.022&  11.365& 0.022&  11.306& 0.018\\ 
			S1589& -10.965 & -2.865 & 0.068 & 0.044&  11.590 & 0.023&  11.352& 0.031&  11.259& 0.019\\ 
			\hline
			\multicolumn{11}{c}{RC stars}\\
			\hline  
			S1074&  -10.953 & -2.98 & 0.063 & 0.045 & 8.650 & 0.018 & 8.122& 0.018&  7.976& 0.018\\ 
			S1084&  -11.002 & -2.881 & 0.069 & 0.053 & 8.619& 0.020  & 8.113& 0.020 &  7.960 & 0.021\\ 
			S1279&   -11.144 & -3.216 & 0.077 & 0.050 & 8.566& 0.024 & 8.072& 0.018&  7.958& 0.024\\ 
			S1316&  -11.181 & -3.157 & 0.115 & 0.069 & 8.618& 0.018 & 8.114& 0.034&  7.996& 0.031\\ 
			S1479&  -11.002 & -3.104 & 0.085 & 0.055  & 8.597& 0.020  & 8.084& 0.024&  7.959& 0.018\\ 
			S1592&  -11.127 & -3.142 & 0.069 & 0.045  & 8.572& 0.021 & 8.087& 0.057&  7.923& 0.023\\ 
			
			\hline
			
		\end{tabular}

	\label{tab:star_param}
\end{table*}

We perform an independent analysis of the chemical composition of 5 of the 10 stars with MyGIsFOS (S984, S792, S1072, S1237, and S1040; see spectra in Fig.~\ref{fig:spec}) with the parameters listed in Table~\ref{tab:param} and obtain the abundances of C, Al, Mg, Si, S, Fe, and Ni (see Table~\ref{tab:abu}). The remaining BSS candidates could not be analysed because they are too hot or rotating too fast (see Table~\ref{tab:candidates}). The stars for which elemental abundances were obtained are circled in Fig.~\ref{fig:iso}.
We select and analyse the chemical composition of two further samples of 10 stars in the TO region and 6 stars on the RC of M67 (see Fig.~\ref{fig:iso}) in order to compare the chemical abundances of the BSS candidates with the abundances found in dwarfs (green squares) and evolved giants (red dots). The results for the abundances are shown in Table~\ref{tab:abu_to} and~\ref{tab:abu_rc}.

\begin{figure*}
	\includegraphics[width=\columnwidth]{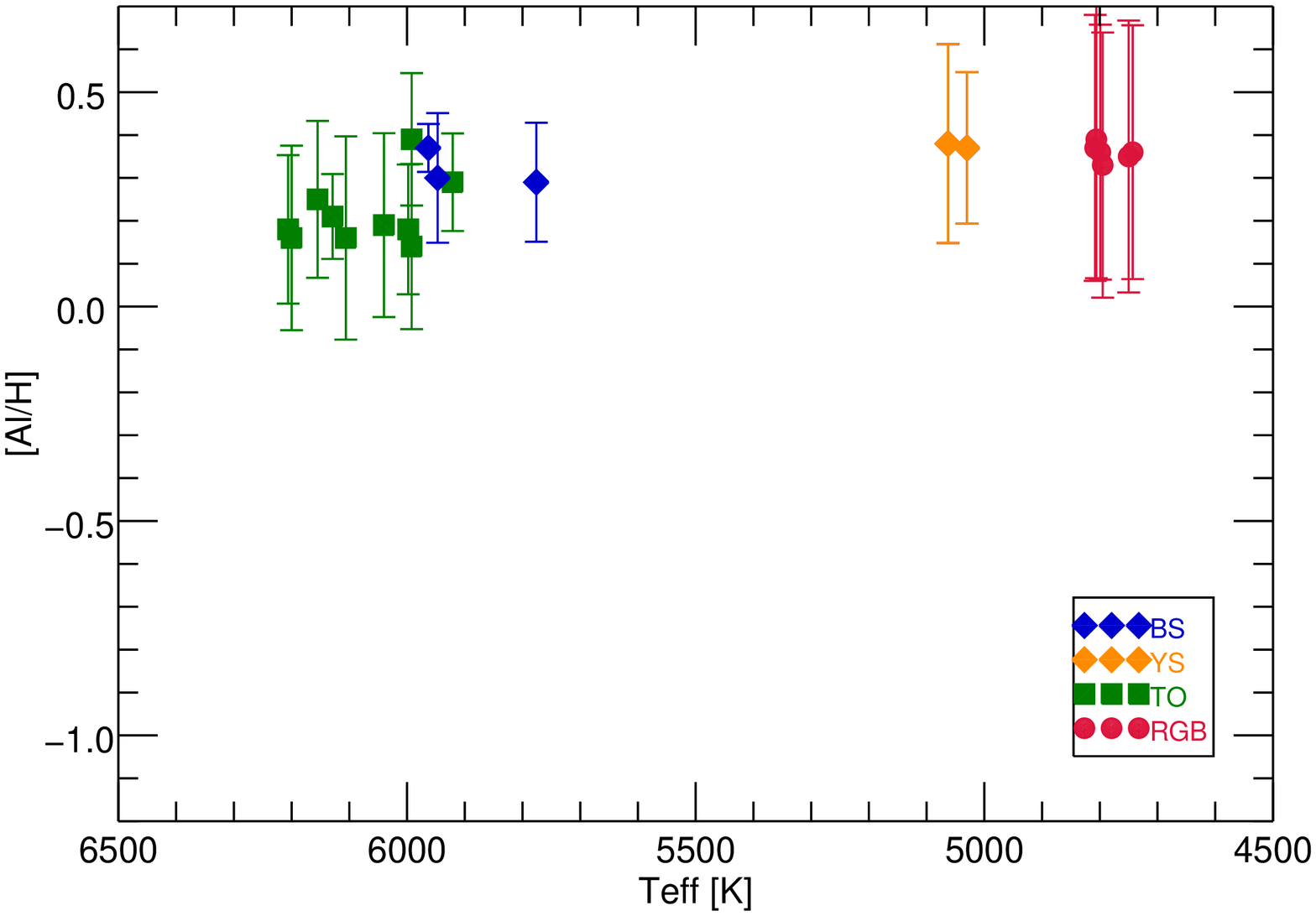}
	\includegraphics[width=\columnwidth]{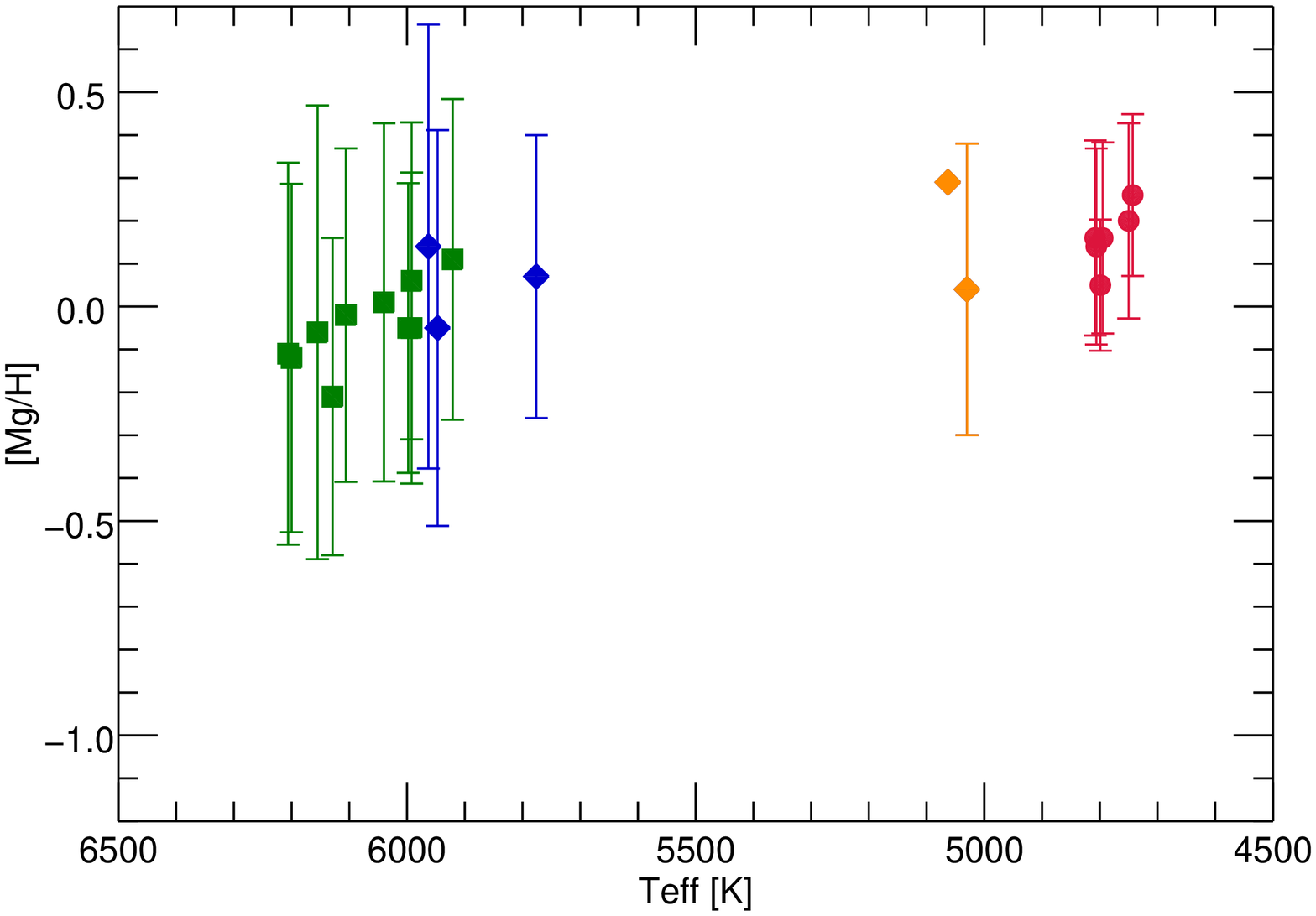}		
	\includegraphics[width=\columnwidth]{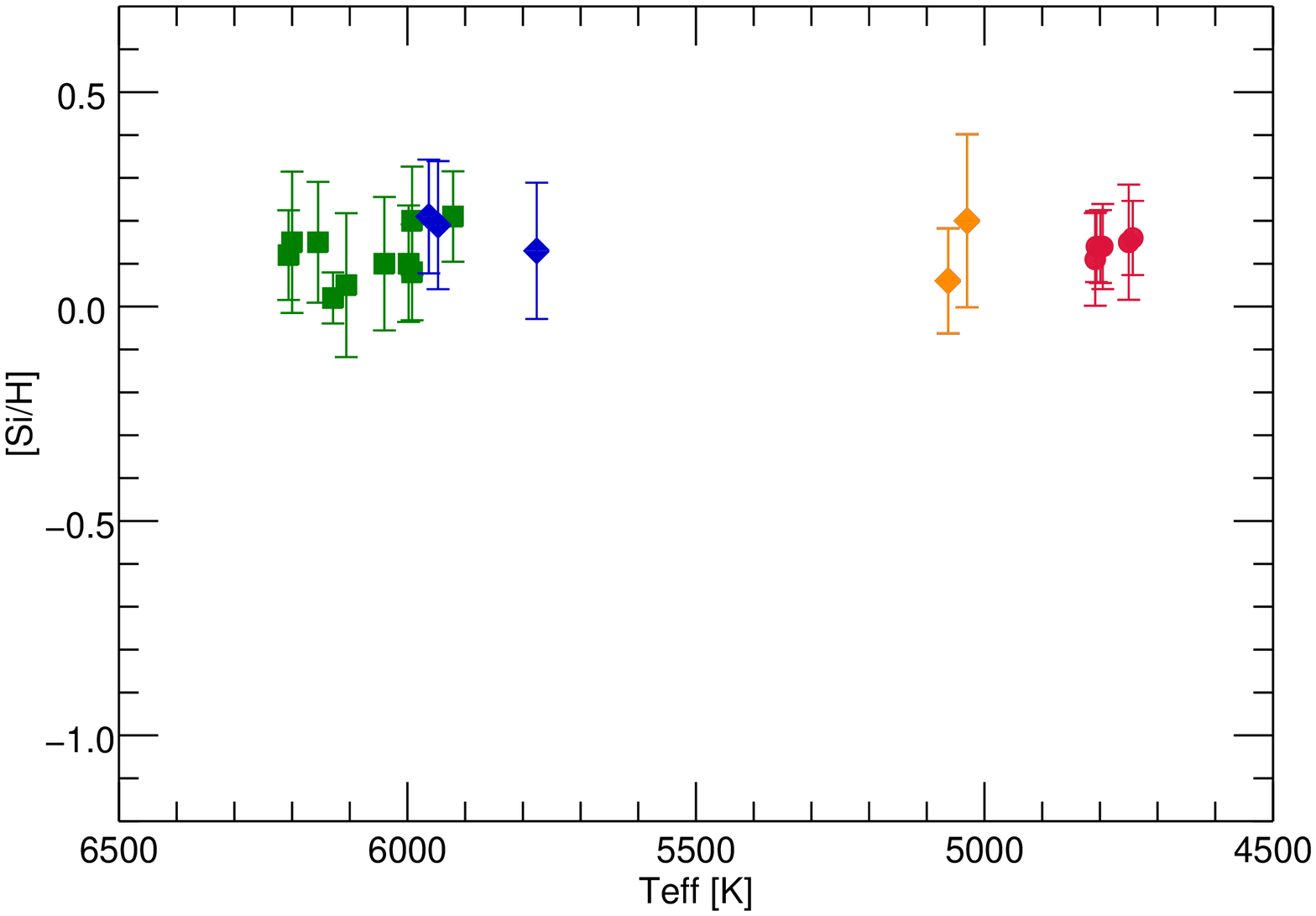}
	\includegraphics[width=\columnwidth]{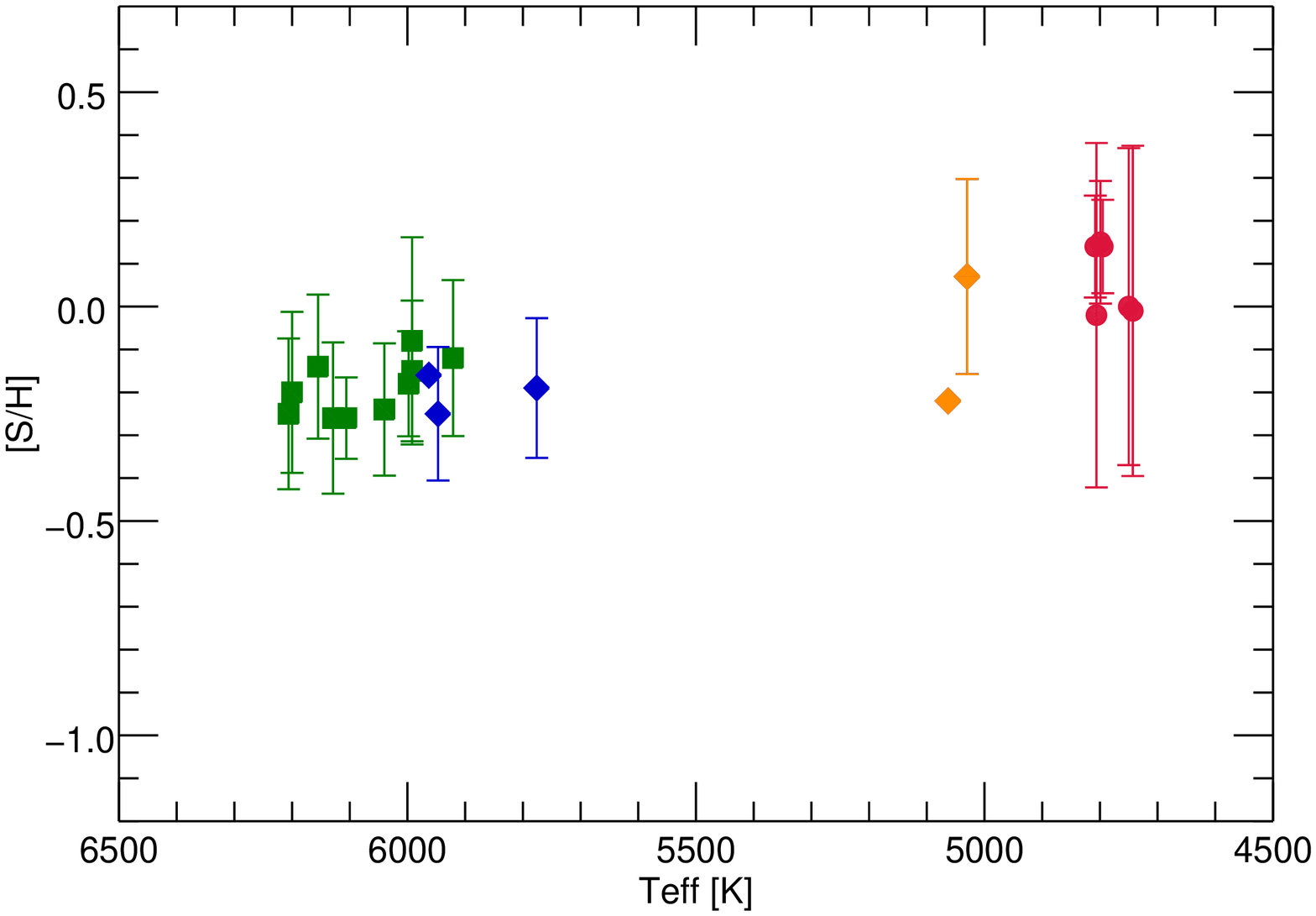}
	\includegraphics[width=\columnwidth]{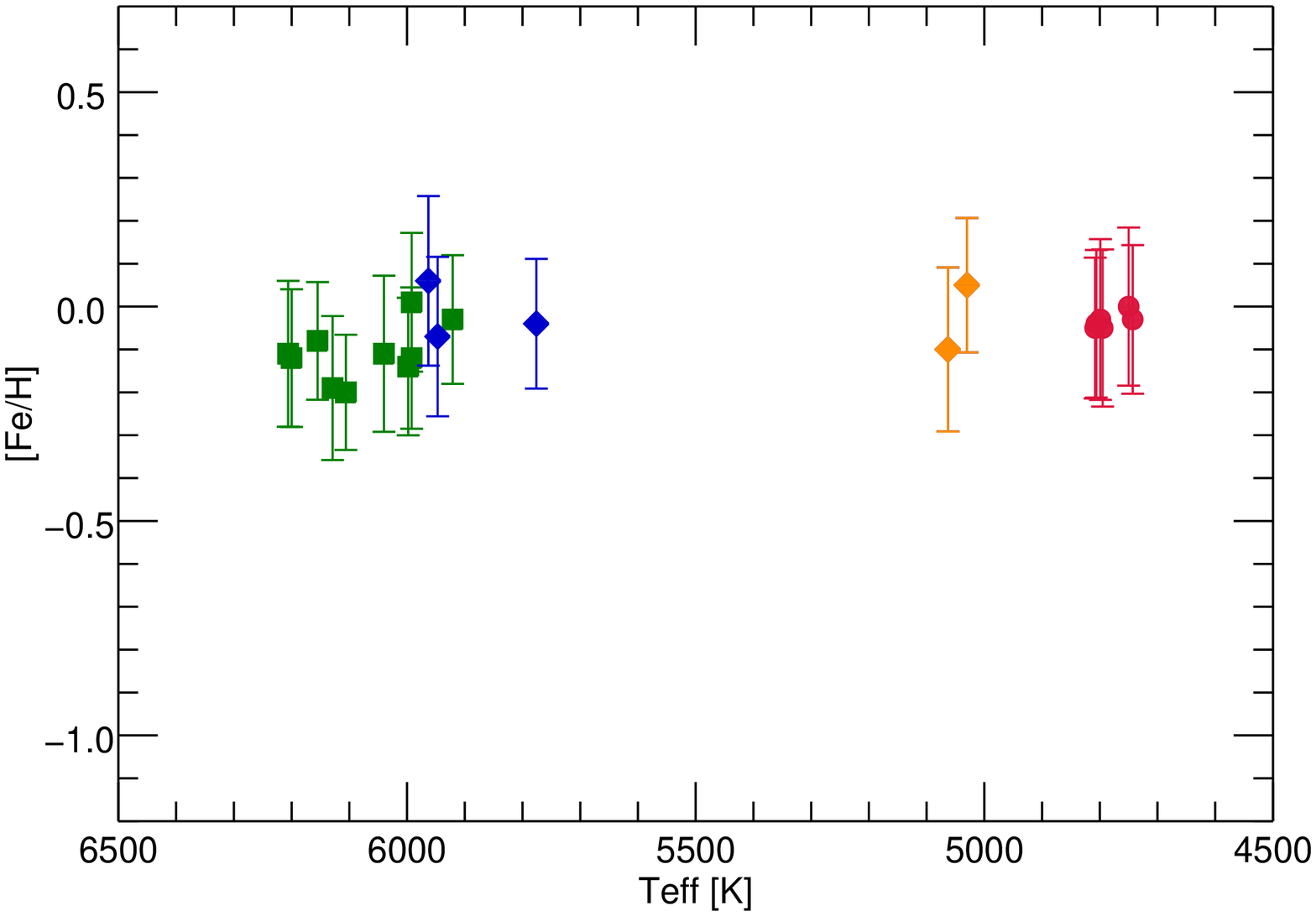}			
	\includegraphics[width=\columnwidth]{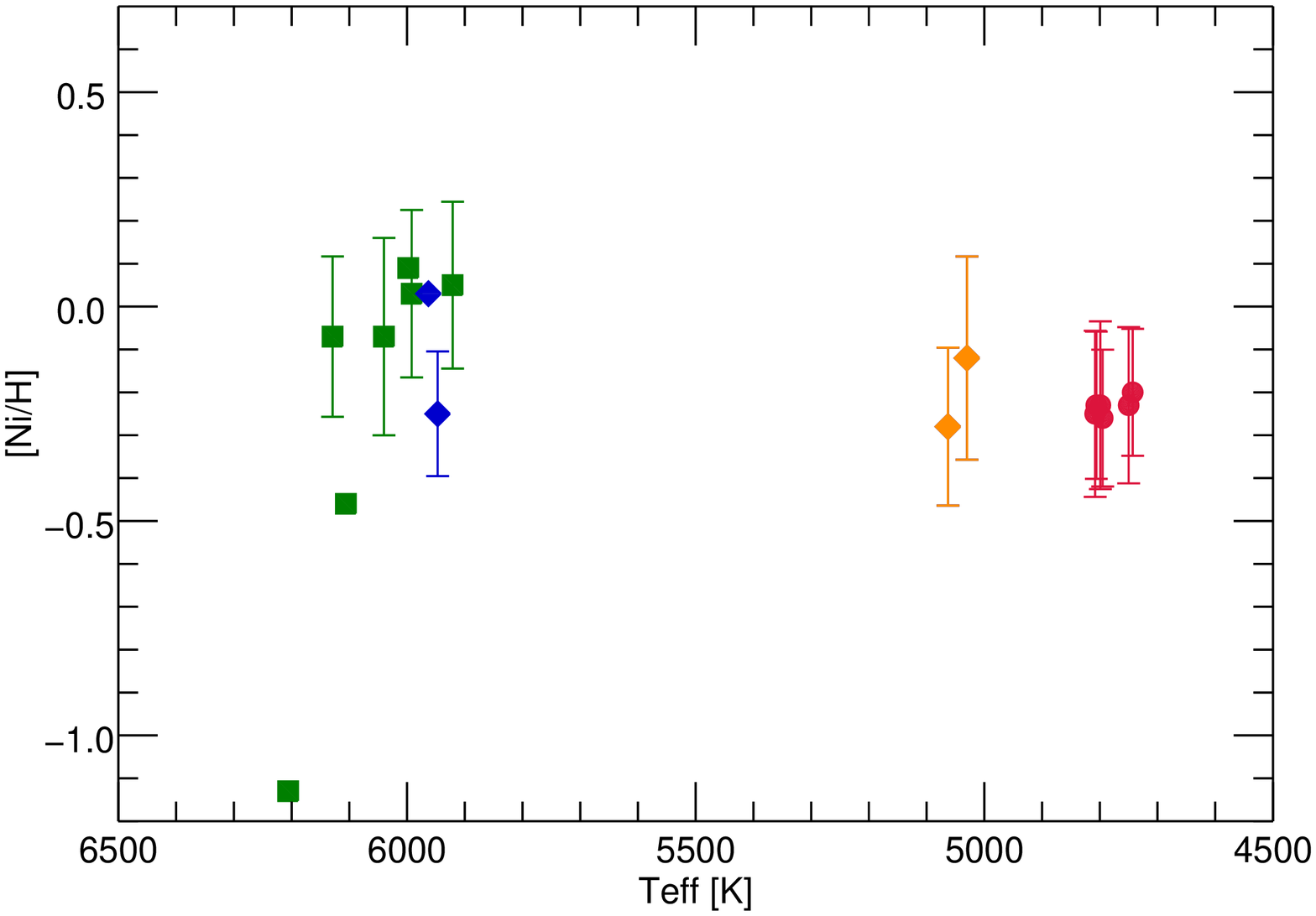}	
	\caption{Abundances derived from our analysis as a function of the APOGEE DR14 calibrated effective temperature $T_\mathrm{eff}$. Colours and symbols are the same as in Fig.~\ref{fig:iso}.}
	\label{fig:abu}
\end{figure*}

For all elements under study, the BSS candidates in M67 present surface abundances in good agreement with the stars in the TO sample (see Fig.~\ref{fig:abu}). To our knowledge, the only previous studies of the chemical composition of BSS candidates in M67 were carried out by \citet{shetrone2000} and \citet{mathys1991}. They performed a chemical analysis of  7 BSS candidates in total and only one of them (S984) is in common with the sample that we analysed with MyGiSFOS. Their results for the C, Mg, Fe, and Ni abundances  are summarised in Table~\ref{tab:abu} and are similar to our findings, showing that the chemical composition of the BSSs candidates is consistent with that of the cluster TO.

\begin{figure}
		\includegraphics[width=\columnwidth]{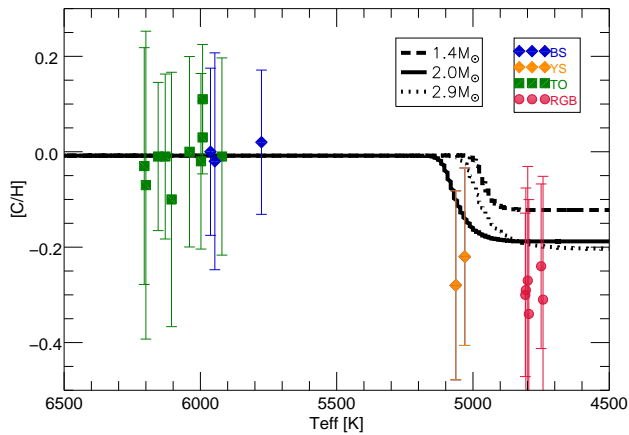}
		\caption{Carbon abundances obtained with our analysis for the BSSs and the control samples. Colours and symbols are the same as in Fig.~\ref{fig:iso}. We also overplot models of stellar evolution calculated for [Fe/H]=0.00 dex and a mass $1.4\,M_{\odot}$ (dashed line), $2.0\,M_{\odot}$ (solid line) and $2.9\,M_{\odot}$ (dotted line).}
		\label{fig:ch}
	\end{figure}

The two evolved BSSs present surface abundances consistent with those of the RC. For most elements these correspond to the abundances at the TO, except for the species that are processed during stellar evolution, such as C.
The carbon abundance of the evolved BSSs is depleted by $\sim 0.25$ dex (see Fig.~\ref{fig:ch}), consistent with the RC abundances derived from our analysis. The carbon abundance is depleted in the RC due to the effect of the first dredge-up (FDU) and possibly of extra-mixing processes (see \citealt{bertelli2017}). The depletion in C abundance of the two evolved BSS could either be explained by their formation history, in the case mass transfer, or by stellar evolution. In fact, these stars are at a later stage of their evolution and should have thus already undergone the FDU  (a detailed discussion on the likelihood of both scenarios for the two objects is presented in Sec.~\ref{sec:disc}). Fig.~\ref{fig:ch} also shows the predicted effect of the FDU with models of mass $1.4\,\text{M}_{\odot}$ (dashed line), $2\,\text{M}_{\odot}$ (solid line), and $2.9\,\text{M}_{\odot}$ (dotted line) with [Fe/H]=0.00 dex from the BaSTI database. The first model represents the mass of the lower RGB in M67, the second that of the star S1237 from the closest isochrone and the third that of S1237 from asteroseismic measurements. We thus expect, following classical stellar evolution, that the two evolved BSSs being more massive would result in a lower C abundance after the FDU than for the stars in the RC.  Nevertheless, the mean [C/H] abundances measured for the evolved BSSs and the RC are consistent with each other and are both lower than the post-FDU abundance for all three models, although all of them are included in the error bars. 
Unfortunately we could not measure the N abundances for the stars under study due to the high temperature and the consequent weakness of the molecular bands from which they are derived.

\begin{figure*}
	\includegraphics[width=17.5cm]{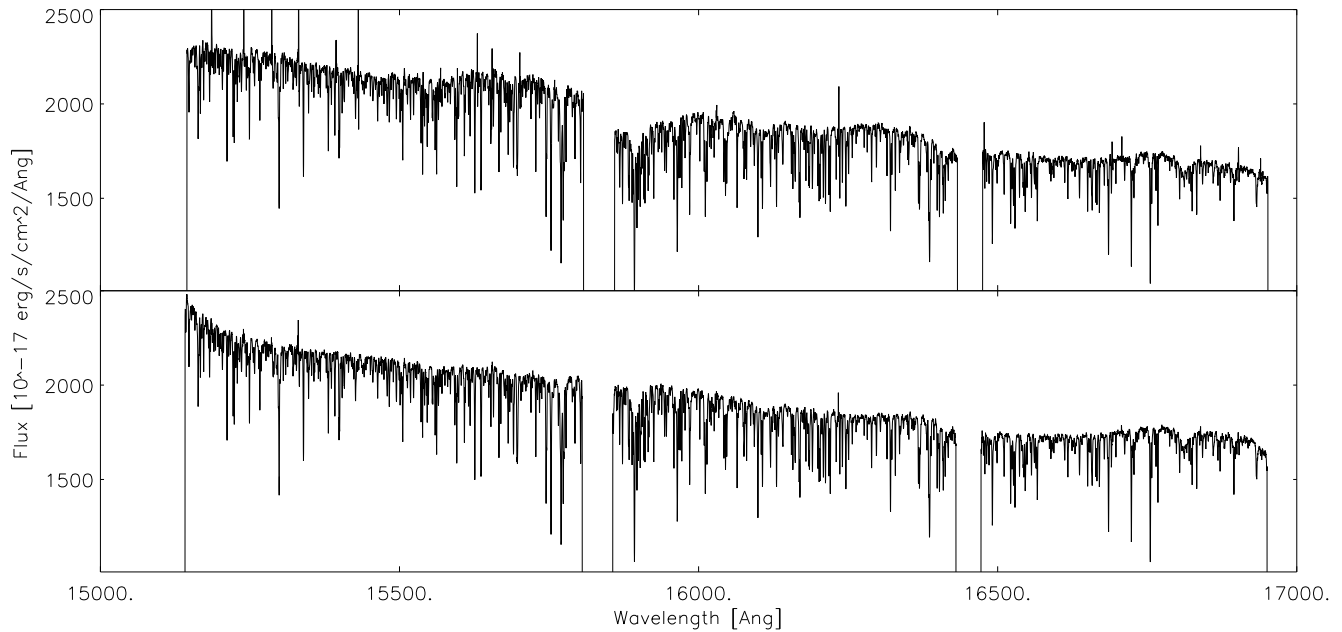}
	\caption{Combination of S1040 spectra with (top panel) and without persistence (bottom panel).}
	\label{fig:per}
\end{figure*}

Some of the stars were observed in several visits by APOGEE and the spectra were then combined before being analysed with ASPCAP. The blue chip of the APOGEE detector is known to suffer from persistence (see, e.g., \citealt{holtzman2015}) and the spectra affected by this problem are flagged in the APOGEE archive. In order to reduce possible errors deriving from the presence of persistence in the spectra under study, we retrieved the spectra of the single visits and combined and analysed only those without persistence, when possible. For S984 and S1237 all visits were affected by high persistence. We therefore tested the influence of persistence on the derivation of the chemical composition in the case of S1040. For this star visits both with and without persistence are available. We separate the spectra with and without persistence and create two combined spectra that we then analysed (see. Fig.~\ref{fig:per}). We found that the presence of persistence in the spectra causes small variations in the abundances derived with our analysis, that are 1 order of magnitude smaller than the uncertainties. We thus conclude that the abundances derived from spectra affected by persistence are reliable, at least in the case of the stars under study.

\begin{figure}
	\includegraphics[width=\columnwidth]{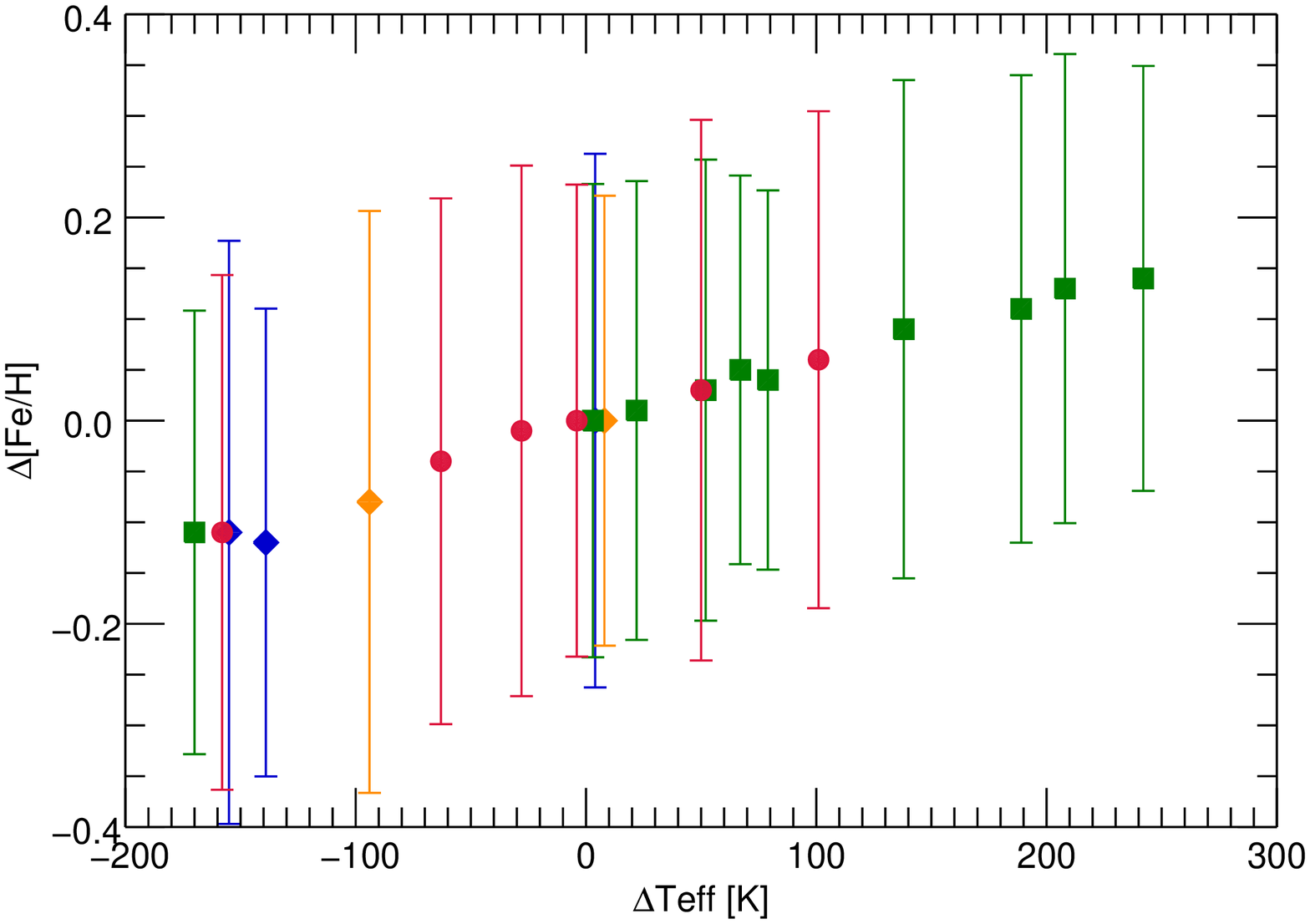}
	\caption{Difference in abundance obtained using two different sets of temperature as a function of the temperature difference. Colours and symbols are as in Fig.~\ref{fig:iso}.}
	\label{fig:teff}
\end{figure}

In addition, we consider two different sets of temperature in order to test the dependence of our results on the input parameters. For one set we adopted the calibrated effective temperature obtained by ASPCAP, while for the second set we calculate the photometric temperature of the stars following the equation of \citet{gonzalez2009}:
\begin{equation}
\theta_\mathrm{eff} =b_0 +b_1X+b_2X^2 +b_3X[\text{Fe/H}]+b_4[\text{Fe/H}]+b_5[\text{Fe/H}]^2,
\end{equation}
where $\theta_\mathrm{eff} =5040/T_\mathrm{eff}$, $X$ is the infrared colour (we use $J-K_s$ corrected for the reddening)\footnote{using $E(B-V)=0.023$ mag \citep{bellini2010} and the extinction law from \citet{cardelli1989}}, and $b_{0..5}$ are the coefficients of the fit. When using $J-K_s$, for dwarf stars the following parameters hold: $b_0=0.6524$, $b_1=0.5813$, $b_2=0.1225$, $b_3=-0.0646$, $b_4=0.0370$, and $b_5=0.0016$, while for giants $b_0=0.6517$, $b_1=0.6312$, $b_2=0.0168$, $b_3=-0.0381$, $b_4=0.0256$, and $b_5=0.0013$.
Both sets of temperatures are listed in Table~\ref{tab:param}. As we show in Fig.~\ref{fig:teff}, there is a trend of the resulting [Fe/H] abundance with temperature, but this effect is smaller than the uncertainty on the abundances and thus should be of no further concern. We also test the effects of variations in $\log g$ and $v_\mathrm{micro}$. Changing, e.g., $\log g$ by $\sim0.25$ dex or $v_\mathrm{micro}$ by $0.04$ km s$^{-1}$ leads to variations in the abundances that are smaller than their uncertainties.

\begin{table*}
	\caption{Input parameters used for the chemical analysis with MyGIsFOS: the calibrated temperatures, gravities, and microturbulences from APOGEE DR14, the photometric temperatures, and the gravities from the isochrones (used when no calibrated gravities from APOGEE DR14 were available). The last column indicates whether the analysed spectra are affected by persistence.}
	\vspace{0.5cm}
	\begin{tabular}{l|c|c|c|c|c|c}
		\hline
		\multicolumn{1}{c|}{Name} &
		\multicolumn{1}{c|}{$T_\mathrm{eff, DR14}$} &
		\multicolumn{1}{c|}{$\log g_\mathrm{DR14}$} &
		\multicolumn{1}{c|}{$v_\mathrm{micro,DR14}$} &
		\multicolumn{1}{c|}{$T_\mathrm{eff, photo}$} &
		\multicolumn{1}{c|}{$\log g_\mathrm{iso}$} &
		\multicolumn{1}{c}{Persistence} \\
		\multicolumn{1}{c|}{} &
		\multicolumn{1}{c|}{[K]} &
		\multicolumn{1}{c|}{[dex]} &
		\multicolumn{1}{c|}{[km s$^{-1}$]} &
		\multicolumn{1}{c|}{[K]} &
		\multicolumn{1}{c|}{[dex]} &
		\multicolumn{1}{c}{} \\
		\hline
		\hline
		S792 & 5947 & --  & 	0.67	&	5943 &	3.65 & N	\\			
		S984 &	5963  &	--   &	0.75 	&	6118 &	3.8	&	Y		\\	
		S1072 &	5776  &	--  &	1.00  	&	5915&	3.4	&N		\\	
		S1237 &	5030 &	3.15 &	1.04 	&	5022 &	2.9	&	Y	\\		
		S1040 &	5063 &	3.03  &	1.12 	&	5157 &	3.2	&	N		\\	
		\hline
		
	\end{tabular}
	\label{tab:param}
\end{table*}

\begin{landscape}

\begin{table}
	\caption{Abundances obtained for the 5 analysed BSSs with MyGiSFOS. When no error on the abundance is listed, the abundance of that element was obtained from a single line and might be therefore less reliable. We also show the abundances obtained by \citet{shetrone2000} and \citet{mathys1991} for 7 BSSs for comparison. The [X/H] abundances from \citet{shetrone2000} and \citet{mathys1991} have been obtained adding [Fe/H] to [X/Fe] and the uncertainties result from error propagation. }
	\vspace{0.5cm}
	
	\begin{tabular}{l|c|c|c|c|c|c|c|c|c|c|c|c|c|c}
		\hline
		\multicolumn{1}{c|}{Name} &
		\multicolumn{1}{c|}{[C/H]} &
		\multicolumn{1}{c|}{e\_[C/H]} &
		\multicolumn{1}{c|}{[Mg/H]} &
		\multicolumn{1}{c|}{e\_[Mg/H]} &
		\multicolumn{1}{c|}{[Al/H]} &
		\multicolumn{1}{c}{e\_[Al/H]} &
			\multicolumn{1}{c|}{[Si/H]} &
			\multicolumn{1}{c|}{e\_[Si/H]} &
			\multicolumn{1}{c|}{[S/H]} &
			\multicolumn{1}{c|}{e\_[S/H]} &
			\multicolumn{1}{c|}{[Fe/H]} &
			\multicolumn{1}{c|}{e\_[Fe/H]} &
			\multicolumn{1}{c|}{[Ni/H]} &
			\multicolumn{1}{c}{e\_[Ni/H]} \\
		\hline
		\hline
	\multicolumn{15}{c}{Stars analysed in our work}\\
		\hline
		S792 & -0.02 &0.23  & 	-0.05	&	0.46 &	0.3 & 0.15 & 0.19 & 0.15 & -0.25 & 0.16 & -0.07 & 0.19 & -0.25 & 0.15 \\		
		S984 &	-0.00  &	0.18   & 0.14	&0.52 &	0.37&	0.06 & 0.21 & 0.13 & -0.16 & -- & 0.06 & 0.2 &0.03 & --	\\	
		S1072 &	0.02 &	0.15   &0.07	&0.33&	0.29	&0.14 &	0.13 & 0.16 &-0.19 &0.16 & -0.04 & 0.15 & --&	--	\\	
		S1237 &	-0.22 &	0.19 &	0.04&	0.34	&	0.37 & 0.18 & 	0.20 & 0.20 & 0.07 & 0.23 & 0.05 & 0.16 & -0.12 & 0.24	\\		
		S1040 &	-0.28 &	0.20  &	0.29	&	-- &	0.38 &0.23 & 0.06 & 0.12 & -0.22 &-- &-0.10 & 0.19 & -0.28 & 0.18		\\	
			\hline
				\multicolumn{15}{c}{Stars from \citet{shetrone2000}}\\
				\hline
			S984 & -0.06 &0.09  & 	-0.01 & 0.07 & -- &-- &-- &--&--&--&0.08	&0.03&0.07 &0.10 \\
			S2204  &0.00 &	0.09   & -0.31 & 0.09 &--&--&--&--&--&--&-0.05	&0.06 &0.12 &0.15 \\
			S975&	-0.07&	0.11 &-0.11& 0.09&--&--&--&--&--&--&0.02&0.06 & -0.23&0.21\\
			S997 &	-0.09&	0.09&-0.10&0.13&--&--&--&--&--&--&	-0.06&	0.03&0.04 &0.10\\
			S1082 &-0.26&	0.07  &	-0.66 & 0.19 &--&--&--&--&--&--&-0.25&	0.05& 0.02&0.11  \\
			\hline
			\multicolumn{15}{c}{Stars from \citet{mathys1991}}\\
			\hline
			S968 &-0.72 & 0.28 & -0.35 & 0.23 &--&--&--&--&--&--&0.03&0.16&0.26 &0.23\\
			S1263	& -0.55 &0.28 & -0.21&0.27&--&--&--&--&--&--&0.09 &0.19&0.42&0.29\\
			\hline
	\end{tabular}

	\label{tab:abu}
\end{table}

\begin{table}
	\caption{Abundances obtained for the 10 TO stars analysed with MyGiSFOS.}
	\vspace{0.5cm}
	
	\begin{tabular}{l|c|c|c|c|c|c|c|c|c|c|c|c|c|c}
		\hline
		\multicolumn{1}{c|}{Name} &
		\multicolumn{1}{c|}{[C/H]} &
		\multicolumn{1}{c|}{e\_[C/H]} &
		\multicolumn{1}{c|}{[Al/H]} &
		\multicolumn{1}{c|}{e\_[Al/H]} &
		\multicolumn{1}{c|}{[Mg/H]} &
		\multicolumn{1}{c}{e\_[Mg/H]} &
			\multicolumn{1}{c|}{[Si/H]} &
			\multicolumn{1}{c|}{e\_[Si/H]} &
			\multicolumn{1}{c|}{[S/H]} &
			\multicolumn{1}{c|}{e\_[S/H]} &
			\multicolumn{1}{c|}{[Fe/H]} &
			\multicolumn{1}{c|}{e\_[Fe/H]} &
			\multicolumn{1}{c|}{[Ni/H]} &
			\multicolumn{1}{c}{e\_[Ni/H]} \\
		\hline
		\hline
		S815& -0.07 &0.32  & 	0.16	&0.22 &	-0.12& 0.41 & 0.15 & 0.17 &-0.2& 0.19 & -0.12 & 0.16&-- &--  \\		
		S598 &0.03  &	0.08  & 0.39	&0.15&	0.06&0.37 & 0.20& 0.13 & -0.08 & 0.24 & 0.01 & 0.16&0.03 & 0.20	\\	
		S756 &-0.01&	0.21   &0.29	&0.11&	0.11	&0.37 & 0.21& 0.11&-0.12 &0.18 & -0.03& 0.15& 0.05&	0.20	\\	
		S1076 &-0.01 &	0.17&	0.21&	0.10	&-0.21 & 0.37 & 0.02 & 0.06& -0.26 & 0.18& -0.19& 0.17& -0.07 & 0.19	\\	
		S1083 &-0.03&	0.25  &	0.18	&0.17	 &-0.11 &0.45 & 0.12 & 0.11 & -0.25 &0.18 &-0.11 & 0.17 & -1.13 & --	\\	
		S1310 & -0.01 & 0.16 & 0.25 & 0.18 & -0.06 & 0.53& 0.15 & 0.14 & -0.14 & 0.17 & -0.08 & 0.14 &-- &-- \\
		S1268 & 0.00 & 0.20 & 0.19 & 0.21 & 0.01 & 0.42 & 	0.10 & 0.16 & -0.24 & 0.15 & -0.11 & 0.18 & -0.07 & 0.23 \\
		S1456 & -0.10 & 0.27 & 0.16 & 0.24 & -0.02 & 0.39 & 0.05 & 0.17 & -0.26 & 0.10 & -0.2 & 0.13 & -0.46 & --\\
		S1429& 0.11  & 0.12 & 0.14 & 0.19 & -0.05 & 0.36 & 0.08 & 0.11 & -0.15 & 0.16 & -0.12 & 0.17 & --& --\\
		S1589 & -0.02 & 0.18 & 0.18 & 0.15 & -0.05 & 0.34& 0.10 & 0.14 & -0.18 & 0.12& -0.14 & 0.16 & 0.09 &--\\
		\hline
		
	\end{tabular}

	\label{tab:abu_to}
\end{table}
\end{landscape}

\begin{landscape}
\begin{table}
	\caption{Abundances obtained for the 6 RC stars analysed with MyGiSFOS.}
	\vspace{0.5cm}
	
	\begin{tabular}{l|c|c|c|c|c|c|c|c|c|c|c|c|c|c}
		\hline
		\multicolumn{1}{c|}{Name} &
		\multicolumn{1}{c|}{[C/H]} &
		\multicolumn{1}{c|}{e\_[C/H]} &
		\multicolumn{1}{c|}{[Al/H]} &
		\multicolumn{1}{c|}{e\_[Al/H]} &
		\multicolumn{1}{c|}{[Mg/H]} &
		\multicolumn{1}{c}{e\_[Mg/H]} &
		\multicolumn{1}{c|}{[Si/H]} &
		\multicolumn{1}{c|}{e\_[Si/H]} &
		\multicolumn{1}{c|}{[S/H]} &
		\multicolumn{1}{c|}{e\_[S/H]} &
		\multicolumn{1}{c|}{[Fe/H]} &
		\multicolumn{1}{c|}{e\_[Fe/H]} &
		\multicolumn{1}{c|}{[Ni/H]} &
		\multicolumn{1}{c}{e\_[Ni/H]} \\
		\hline
		\hline
		S1074& -0.29 &0.21  & 0.39	&0.32 &	0.14& 0.23 & 0.14 & 0.08&-0.02& 0.40 & -0.04 & 0.17& -0.23& 0.17 \\	
		S1084 &-0.27&	0.24& 0.36	&0.30&	0.05&0.15 & 0.14& 0.09 & 0.15& 0.14 & -0.03& 0.19&-0.23 & 0.20	\\	
		S1279&-0.31&0.26  & 0.36&0.30&	0.26	&0.19 & 0.16& 0.09&-0.01 &0.39& -0.03& 0.17& -0.2&	0.15	\\	
		S1316 &-0.34&	0.24&	0.33&0.31	&0.16 & 0.22 & 0.14& 0.10& 0.14& 0.11& -0.05& 0.18& -0.26& 0.16	\\		
		S1479&-0.30&	0.17  &	0.37	&0.31	 &0.16 &0.23 & 0.11& 0.11 & 0.14 &0.12 &-0.05 & 0.16&-0.25 & 0.19	\\	
		S1592&-0.24& 0.17& 0.35& 0.32& 0.2& 0.23 & 0.15& 0.13& 0.0& 0.37& 0.0 & 0.18 & -0.23& 0.18\\
		\hline
		
	\end{tabular}

	\label{tab:abu_rc}
\end{table}

	\begin{figure}
		\includegraphics[width=21cm]{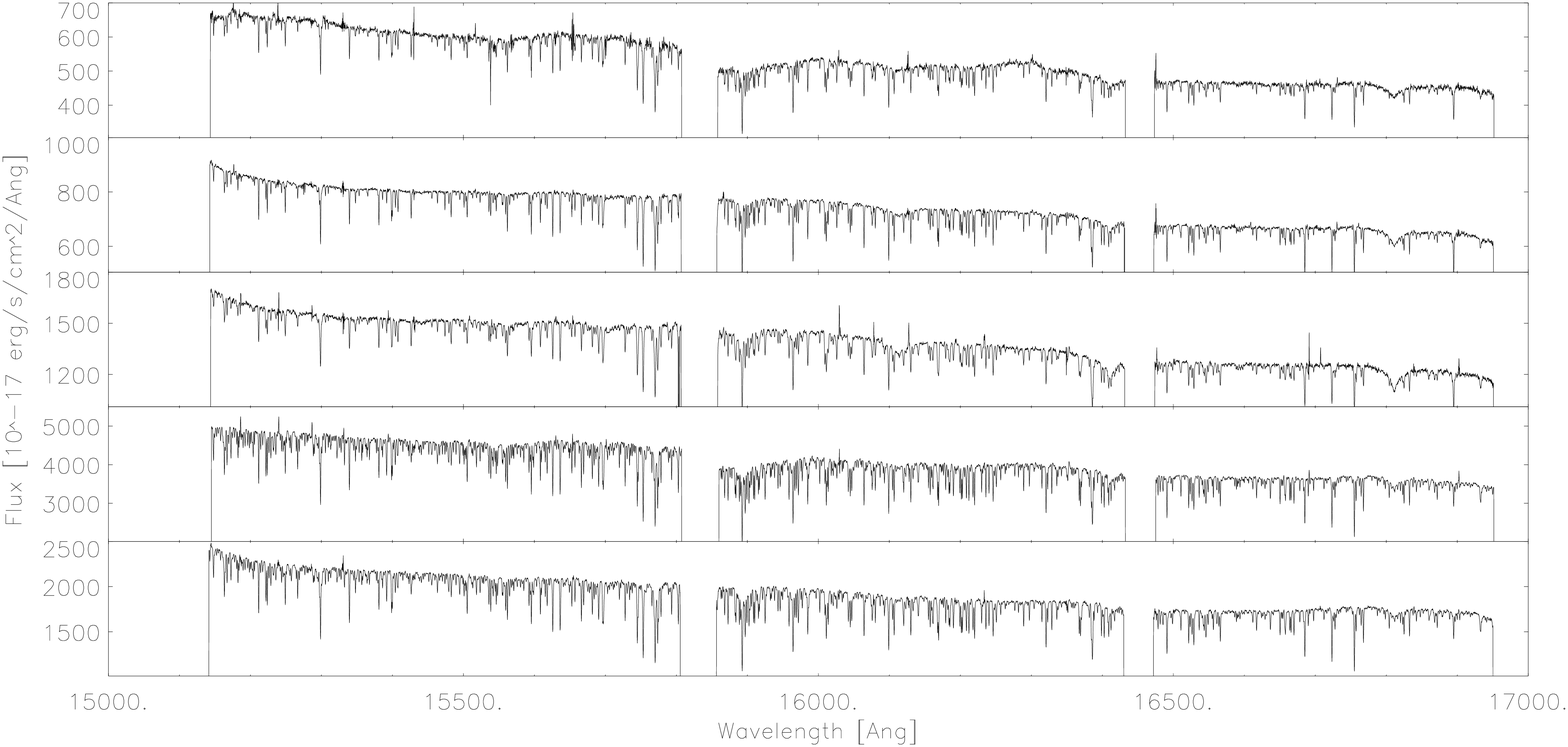}
		\caption{Spectra of the five BSSs analysed in our work. From top to bottom: S984, S792, S1072, S1237, and S1040.}
		\label{fig:spec}
	\end{figure}
\end{landscape}

\section{Discussion}
\label{sec:disc}

In the following we first discuss the general characteristics of BSSs in M67 from APOGEE DR14 data, such as their rotational velocity and spatial distribution, and then summarise the properties of the BSS candidates observed in APOGEE DR14 one by one including our chemical analysis, when available. The properties of the different BSSs are summarised in the Tables~\ref{tab:candidates}, \ref{tab:rv},  \ref{tab:star_param}, \ref{tab:param}, and \ref{tab:abu}.

\subsection{Spatial distribution}

\begin{figure}
	\includegraphics[width=\columnwidth]{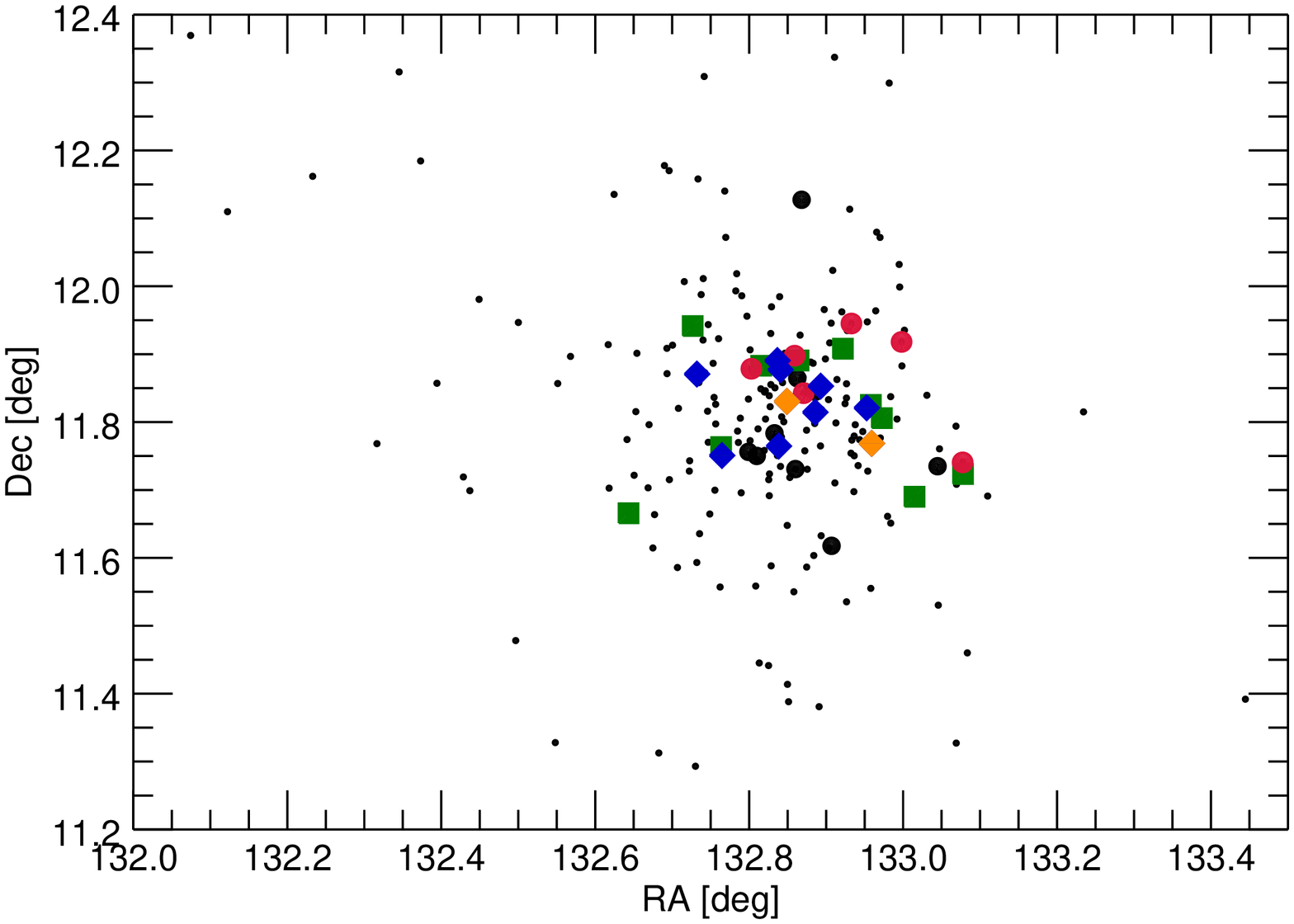}
	\caption{Distribution in equatorial coordinates of the stars considered in this study. Kinematic members of M67 are plotted with small black dots, and the BSS from \citet{geller2015} as large black dots, respectively. BSSs observed in APOGEE DR14, evolved BSSs, TO, and RC stars are displayed with the same colours and symbols as in Fig.~\ref{fig:iso}.}
	\label{fig:space}
\end{figure}

Fig.~\ref{fig:space} shows the spatial position of the sample of 15 BSS candidates from \citet{geller2015} as large black dots. We show stars observed by APOGEE and considered members based on radial velocity and proper motions as small dots.  The spatial distribution of the 8 BSS candidates considered in this work (overplotted as blue diamonds) and of the 2 evolved BSSs (orange diamonds) is concentrated in the central region of the cluster. 

The distribution of the BSSs distances from the centre of the hosting cluster can have different shapes. Globular clusters have been shown to present three different types of spatial distributions (flat, bimodal, or centrally concentrated) that are correlated with the dynamical age of the cluster (see, e.g., \citealt{ferraro1997,ferraro2012}). In the case of open clusters, while BSSs in NGC 188 have been found to have a bimodal spatial distribution  \citep{geller2008}, the density of BSSs in M67, as shown in \citet{geller2015}, decreases rapidly with distance from the cluster centre.  On the other hand, as discussed in \citet{geller2015}, the study of NGC 188 extended to a much larger distance from the cluster centre than their surveys carried out for M67 (extending out to $0.5 \deg$ from the cluster centre). For the authors, it is therefore possible that also for M67 BSSs present a bimodal distribution that has not been discovered yet. \\
We have selected and analysed for membership all stars observed by APOGEE in a $\sim1\deg$ radius around the cluster centre, corresponding roughly to the tidal radius indicated by \citet{kharchenko2013}, and did not find any further BSS candidate. However, APOGEE did not observe every star within this radius and could have potentially missed further BSS candidates at larger distances from the cluster centre.

\subsection{Rotation velocity}
\label{sec:rot}

Another characteristic of BSSs is their (sometimes) high rotational velocity $v\sin i$. Unfortunately, $v\sin i$ does not help us to disentangle the possible formation scenarios, since theory shows that stellar collisions \citep{sills2005}, as well as mass transfer \citep{webbink1976} and coalescence \citep{eggleton2011} can lead to rapid rotation.
The studies carried out on the rotation velocities of BSSs in globular clusters found a correlation between the number of fast rotating BSSs and the density of the cluster: more sparse clusters (e.g. M4, $\omega$ Centauri) have a high concentration of rapid rotating BSSs ($\sim33$\% for $v\sin i>50$ 
$^{-1}$), while in denser clusters (47 Tuc, NGC 6397, M30) the fraction is $\sim4$\% \citep[][and references therein]{ferraro2015}. \citet{simunovic2014} found that in NGC 3201 and NGC 6218 $\sim90\%$ of the BSSs have  $v\sin i$ between 10 and 50 km s$^{-1}$, while $\sim80\%$ of the BSSs in $\omega$ Cen  have $v\sin i$ between 20 and 70 km s$^{-1}$. Interestingly, the fastest rotators tend to be concentrated in the central region for all three clusters.

\begin{figure}
	\includegraphics[width=\columnwidth]{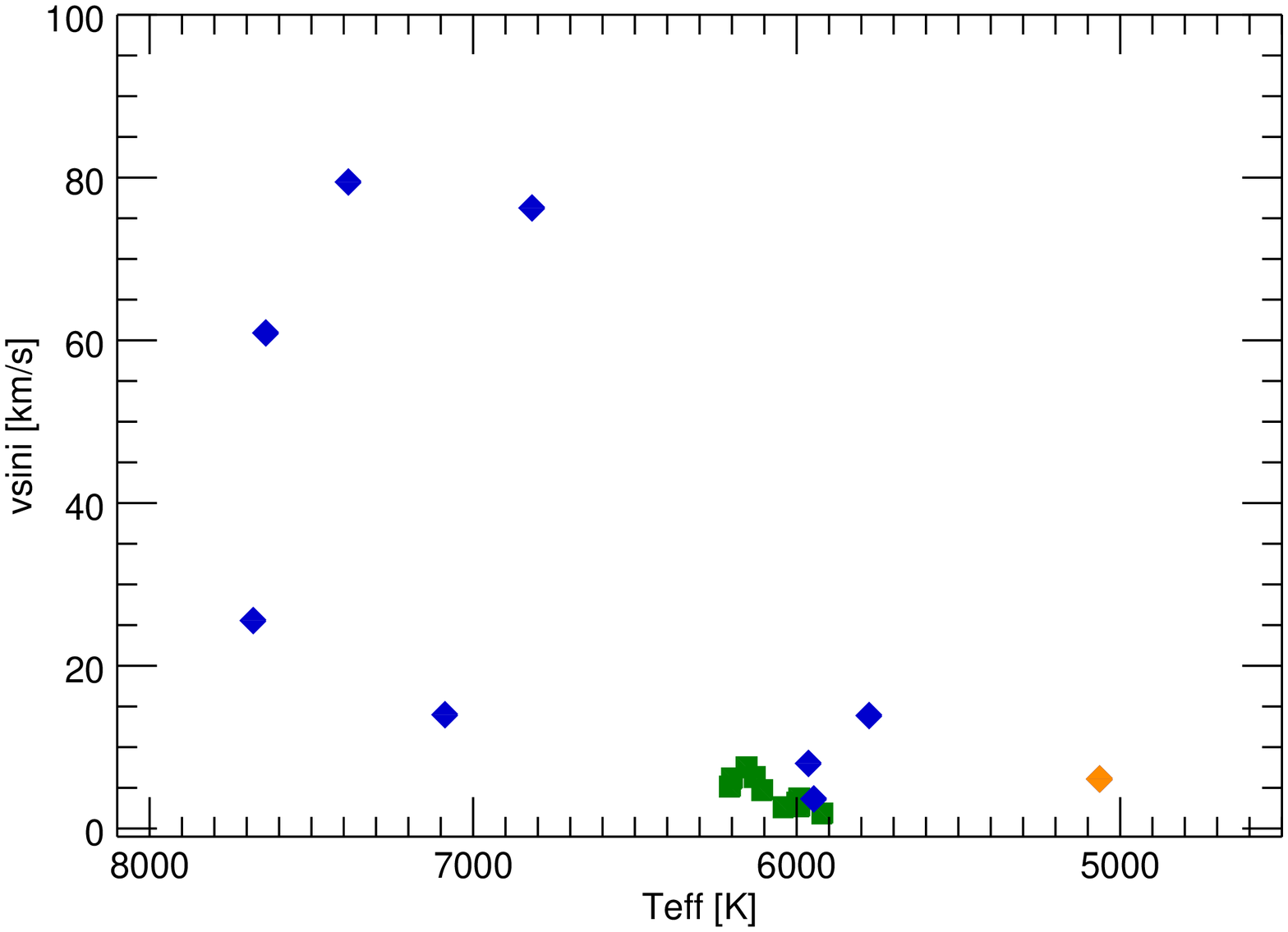}
	\caption{Rotational velocity $v\sin i$ as a function of the uncalibrated ASPCAP effective temperature. Colours and symbols are the same as in Fig.~\ref{fig:iso}.}
	\label{fig:rot}
\end{figure}

One of the outputs of the ASPCAP pipeline is the rotation velocity of the analysed star, $v\sin i$. In the sample of BSS candidates observed by APOGEE DR14, four stars have rotational velocities higher than 20 km s$^{-1}$and up to $\sim 80$ km s$^{-1}$. $v\sin i$ is plotted in Fig.~\ref{fig:rot} as a function of the calibrated effective temperature derived with ASPCAP. All fast rotators have temperatures higher than 6400 K, although there is one star hotter than 7000 K with a rotational velocity lower than 20 km s$^{-1}$. Unfortunately, we could not derive the surface chemical composition of any of these objects due to the line broadening caused by rotation. The stars that we analysed have all $v\sin i < 20$ km s$^{-1}$, and especially S792 and S984 have rotational velocities similar to the TO stars and close to zero. For the RC sample no rotational velocities were measured with ASPCAP. As shown in \citet{mathieu2009} also many BSSs in NGC 188 and NGC 6819 present high rotational velocities, although the highest value that they derive is $\sim50$ km s$^{-1}$. In contrast to the results of \citet{mathieu2009}, we do not find a trend in effective temperature for the measured rotational velocities, although no star cooler than 6400 K is a fast rotator. The presence of a large number of fast rotators in a sparse environment such as M67 is consistent with previous observational evidence whereby fast rotating BSSs are mostly found  in low-density globular clusters \citep{ferraro2015}.

\subsection{The APOGEE sample}

 In the following, we will discuss in more detail the information available from the literature about each of the 10 stars that were subject to the present study, complementing it with the stellar parameters derived by ASPCAP and the results from our analysis. Unless stated differently, the reported radial velocities, rotational velocities and effective temperatures were derived within the APOGEE survey, while the abundances result from our analysis.

\begin{description}
	\item[\textbf{S1267}]  is found to be a binary member and a BSS by \citet{geller2015}, although \citet{balaguer2007} consider its membership uncertain. \citet{sandquist2003b} do not find signs of variability for this star. Nevertheless, S1267 is known to be a spectroscopic binary with a period of 846 d and an eccentricity $e=0.475\pm0.125$ from \citet{latham1996}. The radial velocity $RV=38.7$ km s$^{-1}$ derived with ASPCAP is higher than the cluster mean and the value obtained by \citet{geller2015}, $RV=33.76\pm0.23$ km s$^{-1}$. It is a fast rotator ($vsini=60.9$ km s$^{-1}$) and has $T_\mathrm{eff}=7598$K.\\

\item[\textbf{S1284}] is a $\delta$ Scuti star. Its variable nature is discussed, e.g., in \citet{gilliland1991,sandquist2003b,pribulla2008}. \citet{geller2015} lists it as a binary member of M67 and a BSS. \citet{latham1992} reports a binary period of 4.1828 d and $e=0.266\pm0.045$. Its radial velocity is consistent with the cluster mean and, with $T_\mathrm{eff}=7343$K, it is among the hotter BSSs of the sample. Its rotation velocity is very high, $vsini=79.5$ km s$^{-1}$.\\

\item[\textbf{S1263}] is a single star member of M67 and a BSS according to \citet{geller2015}. \citet{sandquist2003b} state that it might be a variable star with low amplitude, while \citet{pribulla2008} do not find any sign of variability for this object. It has a radial velocity consistent with the cluster, it seems to rotate much slower than the BSSs discussed so far ($vsini=25$ km s$^{-1}$), and it has  $T_\mathrm{eff}=7599$K. S1263 is one of the BSSs analysed by \citet{mathys1991} who found it to be depleted in both C and O but with an iron abundance consistent with that of the cluster ($\mathrm{[Fe/H]}=0.09\pm0.19$ dex).\\

\item[\textbf{S1082}] is the best studied BSS in M67. It is known to be a triple system composed of a close binary with a BSS and a TO star and a third component, probably a BSS itself. The spectroscopic study of \citet{vandenberg2001} revealed the presence of three components in the system. X-ray emission found by \citet{belloni1998} suggests that the close system might be a magnetically active RS CVn-type eclipsing binary. In \citet{sandquist2003a}, the light curve of this stellar system was studied in detail and the authors confirmed S1082 to be formed by a close binary with an orbital period of 1.08 d and a second component which is either a spatial superposition, or a third star dynamically bound but on a large eccentricity orbit with period of 1189 d. \citet{pribulla2008} further studied the asymmetry of the  light-curve of S1082, concluding that it is probably caused by photospheric spots in the non-BSS companion in the close binary. \citet{leigh2011} developed an analytic method based on energy conservation in stellar encounters to determine the formation scenario of multiple stellar systems. They analysed the specific case of S1082 and came to the conclusion that, if the close binary and the second component are indeed dynamically connected, they must have formed through a single 3+3 encounter. Since such encounters are very rare considering the present density of M67, they conclude that the second component of S1082 is probably only spatially overlapping the close binary. If, on the other hand, one takes into consideration that M67 might have lost part of its mass due to evaporation and interaction with the Galactic disc and that its density might have been a factor of 2 higher in the past, then the probability for the encounter of two triple systems is much higher. S1082 was part of the study of \citet{shetrone2000}, who found its abundances to be on average 0.2 dex lower than the cluster mean. The effective temperature derived in APOGEE DR14 is $T_\mathrm{eff}=7087$K, much higher than its photometric temperature $T_\mathrm{eff}=6318$K, but it is not a fast rotator ($vsini=13.99$ km s$^{-1}$).  \\

\item[\textbf{S752}] is considered an Am star with a possible flare detected by \citet{sandquist2003b}. It was found to be a long-period binary with 1003 d and $e=0.317\pm0.123$ by \citet{latham1996}. \citet{geller2015} list it as a binary member and a BSS. Its radial velocity is not consistent with the cluster mean ($RV=25.68\pm0.14$ km s$^{-1}$) and lower than the value derived by \citet{geller2015} ($RV=31.29\pm0.35$ km s$^{-1}$), which could be due to the binary nature of this object. It is a fast rotator ($vsini=77$ km s$^{-1}$), and has $T_\mathrm{eff}=6631$K.\\

\item[\textbf{S792} and \textbf{S984}] are considered single members of M67 by \citet{geller2015}, although they comment that these two stars might possibly be binaries with a very long period that could not be detected in their survey. Similarly, \citet{sandquist2003b} do not find any variability in the light curve of these two stars. \citet{geller2015} argue that these two stars are not BSSs since their position on the CMD is too close to the TO, in a region that is expected to be populated by binaries consisting of MS stars. Indeed, Fig.~\ref{fig:iso} shows how they are included in the area between the single-star isochrone of M67 and the equal-mass binary isochrone, obtained by shifting the former by -0.75 mag.  Nevertheless, this notion alone does not exclude the possibility that these stars are BSSs (single or in a long-period binary) formed from stars with masses below the TO.\\
\citet{shetrone2000} found the  Li abundance of S984 to be consistent with that of the TO. This is in disagreement with our current understanding of all BSS formation channels, from which we would expect to observe a large depletion of Li, and seems to indicate that S984 is indeed not a BSS. \citet{shetrone2000} combined their derived RV with other data from \citet{mathieu1986} and inferred a possible orbital period of $\sim1.5$d for this star. They argue, however, that if the system was indeed a close binary it should be tidally locked, but in that case its measured lithium abundance would be too low. Unfortunately, Li measurements are not available for S972, since this star was not analysed by \citet{shetrone2000} and there are no strong Li-features in the spectral range covered by APOGEE.

The three visits available in APOGEE for S984 unveil a trend of decreasing heliocentric velocity (34.14, 33.47, and 32.77 km s$^{-1}$, respectively) in a period of 5 months. Although this is not enough to infer a binary period for this star and the number of visits is too low for any final statement, the maximal difference in heliocentric velocity (1.37 km s$^{-1}$) is two orders of magnitude larger than the median error on the single measurements (0.04 km s$^{-1}$). No such trend is visible in the eight visits available for S792 in a period of four months. The APOGEE rotational velocities of both stars are low, $vsini=3.6$ km s$^{-1}$ and  $vsini=8$ km s$^{-1}$ respectively, consistent with the rotation velocities measured for TO stars. Their ASPCAP temperatures ($T_\mathrm{eff}=5910$ K and $T_\mathrm{eff}=5927$K) are also consistent with the TO, as are the abundances that we obtain in our analysis. In summary, given its Li-abundance \citep[see][]{shetrone2000}, S984 is unlikely to be a BSS. Its position above the TO could be explained if the star would be part of a binary system which, given the available APOGEE data, cannot be excluded or confirmed. S792, as suggested by  \citet{geller2015}, can either be a binary with a long period that could not yet be detected and be composed of TO stars or a BSS formed through collision, since its chemical composition does not exclude this scenario.\\

\item[\textbf{S1072}] is listed as a spectroscopic binary member and a yellow giant of M67 by \citet{geller2015}. \citet{sandquist2003b} did not find any variation in its light curve. \citet{liu2008} analysed S1072 and found it to have $\log g$ lower than 4 dex. Its peculiar position above the SGB and its colour being much redder than the other BSSs in the sample could indicate that S1072 is a BSS in the process of evolving to a giant. \citet{mathieu1990} discovered that it is a binary with a 1495 d period and $e=0.32\pm0.07$. \citet{vandenberg2002} found that S1072, as well as S1040 and S1237 (see below), present an unexplained X-ray emission. The radial velocity of S1072 is consistent with the cluster's velocity distribution and it has a low rotational velocity ($vsini=13.88$ km s$^{-1}$). The abundances that we derive are consistent with those of the TO, and also C is not depleted. The absence of carbon depletion would rule out a formation through mass transfer, although as \citet{ferraro2006} point out, the C depletion expected in BSSs with a mass transfer history might be a transient one and, if indeed S1072 is already evolving along its SGB, the mass-transfer event would have happened in the non-recent past. If S1072 is a long-period and high-eccentricity binary composed of a BSS and a further companion, the original system is likely to have been formed by at least 3 stars. We therefore suggest that S1072 might have formed through Kozai cycles in a hierarchical triple system \citep{perets2009}.\\

\item[\textbf{S1040}] is most likely an evolved BSS, and lies bluewards of the M67 RGB. \citet{mathieu1990} found this object to be a binary with period 42.8 d.  \citet{landsman1997} carried out a detailed study of this object based on spectroscopic and photometric data and identified it as a binary system with a He white dwarf as a secondary companion. This indicates that part of the mass of the He WD was stripped from the star before it could reach the He burning phase and was accreted onto the primary companion. Thus, S1040 can be considered a typical example of an evolved BSS formed through mass transfer in a binary system. \citet{sandquist2003b} find a drop in luminosity corresponding to the passage of the white dwarf in front of the giant, but they comment that this variability cannot be caused by an eclipse and it is probably due to other effects due to the passage of the WD. We find that S1040 has abundances consistent with those of the RC stars. This also includes C, which is found to be depleted by $\sim0.25$ dex with respect to the TO. This could be the C depletion expected from mass-transfer BSSs, although it would mean that the signature has remained visible for a long time after the accretion event, given the evolved nature of S1040. Another possibility is that we are observing the effects of classical stellar evolution taking place after the formation of the BSS, and in particular the FDU.\\

\item[\textbf{S1237}] is an evolved BSS situated between the SGB and the RC of M67, with a slightly bluer colour than the RGB (see Fig.~\ref{fig:iso}) and $\sim1$ mag more luminous in K$_s$ than S1040. This star has been studied in detail by \citet{leiner2016} who, based on asteroseismological  data from the Kepler K2 mission \citep{howell2014}, found S1237 to be a binary system whose primary has $M=2.97\pm0.24\,M_{\odot}$ and $R=9.27\pm0.19\,R_{\odot}$. Considering that the TO mass of M67 is $\sim1.3\,M_{\odot}$, these results seem to indicate that S1237 had an uncommon evolution history and is most likely an evolved blue straggler star. Despite the fact that the spectrum of S1237, similarly to other MS stars of M67, shows a FUV excess, SED fitting indicates that the secondary companion is probably an upper-MS star or a BSS close to the TO of M67 \citep[see][]{leiner2016}. As for the formation history of S1237, \citet{leiner2016} suggest that a collision scenario as well as a Kozai-cycle-induced merger are possible mechanisms. A mass-transfer scenario seems unlikely given the asteroseismic mass of the object and the apparent absence of a white dwarf companion. Similarly to S1040, S1237 is depleted in C consistent with the RC. If for this object a mass-transfer scenario is unlikely, we can only explain the carbon depletion with stellar evolution.

\end{description}

\section{Summary}
\label{sec:concl}

We analyse the chemical composition of three candidate BSSs as well as of two known evolved BSSs in the old open cluster M67 and we discuss the results together with information gathered from the literature. We find that BSS candidates share the same surface abundances as TO stars. Based on its position in the CMD and on past studies on its variability and chemical composition, the BSS candidate  S984 is likely to be a long-period binary whose brightness variability could not yet be detected, particularly given that its high Li abundance \citep{shetrone2000} is inconsistent with the expectations for a BSS. Following a similar line of reasoning, S792 is either a long-period binary formed by two TO stars or a single BSS. If the second is true, the measured abundances, similar to those of the TO stars, hint at a collisional formation scenario. 

S1072 also shares the same chemical composition as the dwarf stars of M67 and presents no C depletion. A possible explanation for the absence of a chemical signature indicating mass transfer can be found in \citet{ferraro2006}, where the authors suggest that the signature might be visible only for a short time, before the mixing in the stellar interior averages it out. The red colour of this star has been interpreted already in the past as a sign that S1072 might be in the process of evolving along its subgiant branch. In the case of a mass transfer history, this event would not have taken place in the recent past and the corresponding CO signature would have likely already disappeared. At the same time, if S1072 indeed is a BSS in binary with an orbital period of 1495 d and eccentricity of 0.32, it is unlikely to have undergone mass transfer with the present companion. We suggest that S1072 could have formed through a Kozai-cycle-induced merger in a hierarchical triple system.  

S1040 is known to be an evolved BSS with a He-core companion and thus it most likely formed through mass transfer. We measure a depletion in C but, if this is the signature of a mass-transfer event and such signatures are transient as suggested by \citet{ferraro2006}, it would be strange to observe it in a star that has already evolved so far along the RGB. On the other hand the C depletion is higher (although consistent within the errors and with the RC abundances) than the theoretical predictions of the FDU for stars in the mass range $1.4-2.9\,M_{\odot}$. 

A similar line of thoughts holds for S1237, except that for this object \citet{leiner2016} have shown that a formation scenario through mass transfer is unlikely in comparison to the possibility of a collision or Kozai-cycle induced merger in the context of a dynamical encounter between multiple systems. Thus, in this case the C depletion that we measure must be due to evolutionary effects rather than formation processes.

In summary, we find a depletion in carbon only for the evolved BSSs known in M67, S1040, and S1237, but we argue that this is most likely due to processes of stellar evolution rather than to a mass transfer history, although measurements of the nitrogen abundances of these stars would be necessary to shed more light on the matter. Also, the errors on the C abundances do not yet allow us to distinguish between the different amount of C depletion suffered by stars of different mass after the FDU. Our results confirm the findings of globular cluster studies, where very few carbon-depleted BSSs are found. [C/H] estimations in the hotter BSSs would be necessary in order to obtain a complete picture and an estimate of the significance of mass transfer in the formation of BSSs.

\section*{Acknowledgements}

This work was supported by Sonderforschungsbereich SFB 881 "The Milky Way System" (subproject B5) of the German Research Foundation (DFG). The authors thank Maurizio Salaris for providing the models shown in Fig.~\ref{fig:ch}. CBM thanks Giacomo Beccari, Francesco Ferraro, and Gergely Hajdu for fruitful discussions. The authors thank the anonymous referee for the constructive feedback that helped improving the present work.

Funding for the Sloan Digital Sky Survey IV has been provided by the Alfred P. Sloan Foundation, the U.S. Department of Energy Office of Science, and the Participating Institutions. SDSS acknowledges support and resources from the Center for High-Performance Computing at the University of Utah. The SDSS web site is www.sdss.org.

SDSS is managed by the Astrophysical Research Consortium for the Participating Institutions of the SDSS Collaboration including the Brazilian Participation Group, the Carnegie Institution for Science, Carnegie Mellon University, the Chilean Participation Group, the French Participation Group, Harvard-Smithsonian Center for Astrophysics, Instituto de Astrof\'isica de Canarias, The Johns Hopkins University, Kavli Institute for the Physics and Mathematics of the Universe (IPMU) / University of Tokyo, Lawrence Berkeley National Laboratory, Leibniz Institut f\"ur Astrophysik Potsdam (AIP), Max-Planck-Institut f\"ur Astronomie (MPIA Heidelberg), Max-Planck-Institut f\"ur Astrophysik (MPA Garching), Max-Planck-Institut f\"ur Extraterrestrische Physik (MPE), National Astronomical Observatories of China, New Mexico State University, New York University, University of Notre Dame, Observat\'orio Nacional / MCTI, The Ohio State University, Pennsylvania State University, Shanghai Astronomical Observatory, United Kingdom Participation Group, Universidad Nacional Aut\'onoma de M\'exico, University of Arizona, University of Colorado Boulder, University of Oxford, University of Portsmouth, University of Utah, University of Virginia, University of Washington, University of Wisconsin, Vanderbilt University, and Yale University.

This work has made use of data from the European Space Agency (ESA) mission
{\it Gaia} (\url{https://www.cosmos.esa.int/gaia}), processed by the {\it Gaia}
Data Processing and Analysis Consortium (DPAC,
\url{https://www.cosmos.esa.int/web/gaia/dpac/consortium}). Funding for the DPAC
has been provided by national institutions, in particular the institutions
participating in the {\it Gaia} Multilateral Agreement.

%%%%%%%%%%%%%%%%%%%%%%%%%%%%%%%%%%%%%%%%%%%%%%%%%%



\bibliographystyle{mnras}
\bibliography{Bibliography} % if your bibtex file is called example.bib

%\begin{thebibliography}{99}
%\end{thebibliography}

%%%%%%%%%%%%%%%%%%%%%%%%%%%%%%%%%%%%%%%%%%%%%%%%%%

%%%%%%%%%%%%%%%%% APPENDICES %%%%%%%%%%%%%%%%%%%%%

%\appendix


%%%%%%%%%%%%%%%%%%%%%%%%%%%%%%%%%%%%%%%%%%%%%%%%%%

% Don't change these lines
\bsp	% typesetting comment
\label{lastpage}
\end{document}